\documentclass[10pt,letterpaper,twoside]{amsart}

\usepackage{xcolor}

\definecolor{astral}{RGB}{0,164,239}

\usepackage{fancyhdr}

\usepackage{titlesec}

\titleformat{\section}
{\normalfont\fontfamily{phv}\fontsize{12}{15}\bfseries\filcenter\scshape\color{astral}}{\thesection}{1em}{}

\titleformat{\subsection}
{\normalfont\fontfamily{phv}\fontsize{10}{13}\bfseries\color{astral}}{\thesubsection}{1em}{}

\titleformat{\subsubsection}
{\normalfont\fontfamily{phv}\fontsize{10}{13}\selectfont\color{astral}}{\thesubsubsection}{1em}{}

\usepackage[colorlinks=true]{hyperref}

\definecolor{ccol}{RGB}{246, 83, 20}
\hypersetup{colorlinks,citecolor={ccol},linkcolor={ccol}, urlcolor={ccol}}

\usepackage{amsmath,amsfonts,amsbsy,amsgen,amscd,mathrsfs,amssymb,amsthm}

\usepackage[cal=cm]{mathalfa}

\usepackage{url}

\usepackage{mathtools}

\usepackage[font=small,margin=0.25in,labelfont={sc},labelsep={colon}]{caption}

\usepackage{microtype}
\usepackage{enumitem}

\usepackage{amsmath,amsfonts,bm}

\usepackage{xcolor}
\usepackage{colortbl}

\def\eqref#1{equation~\ref{#1}}
\def\1{\bm{1}}

\DeclareMathAlphabet{\mathsfit}{\encodingdefault}{\sfdefault}{m}{sl}
\SetMathAlphabet{\mathsfit}{bold}{\encodingdefault}{\sfdefault}{bx}{n}

\DeclareMathOperator*{\argmin}{arg\,min}

\DeclareMathOperator{\sign}{sign}
\DeclareMathOperator{\clip}{clip}

\usepackage{setspace}

\usepackage{graphicx}
\usepackage{booktabs,longtable,tabu} 
\usepackage{multirow} 
\usepackage{float}

\usepackage[full]{textcomp}

\usepackage{bm} 
\usepackage[utf8]{inputenc} % allow utf-8 input
\usepackage[T1]{fontenc}    % use 8-bit T1 fonts
\usepackage{dcolumn}
\usepackage{nicefrac}       % compact symbols for 1/2, etc.
\usepackage{microtype}      % microtypography

\usepackage{overpic}

\usepackage{upgreek}
\usepackage{bm}
\usepackage{mathrsfs}
\usepackage{afterpage}

\usepackage{subcaption}
\usepackage{multirow}
\usepackage{graphicx}
\usepackage{epstopdf}
\usepackage{algorithmic}

\usepackage{pgfplots}
\usepackage{mathtools}
\usepackage{overpic}
\usepackage{tikz}
\usetikzlibrary{matrix,positioning,shapes,shadows,arrows,calc,decorations.pathreplacing}
\usetikzlibrary{arrows.meta}

\definecolor{darkred}{RGB}{228,26,28}
\definecolor{darkblue}{RGB}{44,127,184}
\definecolor{magentaCB}{RGB}{221,28,119}

\definecolor{morange}{RGB}{255, 187, 0}
\definecolor{mblue}{RGB}{ 0, 161, 241}
\definecolor{mgreen}{RGB}{124, 187, 0}
\definecolor{mred}{RGB}{246, 83, 20}

\definecolor{lightred}{RGB}{255, 235, 234}
\definecolor{graytable}{RGB}{240,240,240}

\usepackage{xspace}

\usepackage[utf8]{inputenc} % allow utf-8 input
\usepackage{amsmath}

\title[JumpReLU]{JumpReLU: A Retrofit Defense Strategy for Adversarial Attacks}

\author[Erichson, Yao, and Mahoney]{N. Benjamin Erichson$^{1*}$, Zhewei Yao$^{2*}$, and Michael W. Mahoney$^{1}$}
\thanks{$^{1}$ ICSI and Department of Statistics, University of California at Berkeley, Berkeley, CA 94720.}

\thanks{$^{2}$ Department of Mathematics, University of California at Berkeley, Berkeley, CA 94720.}

\thanks{$^{*}$ Indicates equal contributions. Corresponding author: N. Benjamin Erichson (\href{mailto:erichson@berkeley.edu}{erichson@berkeley.edu}).}

\date{}


\makeatletter
\DeclareRobustCommand\onedot{\futurelet\@let@token\@onedot}
\def\@onedot{\ifx\@let@token.\else.\null\fi\xspace}

\def\eg{\emph{e.g}\onedot} 
\def\ie{\emph{i.e}\onedot}

\def\etal{\emph{et al}\onedot}
\makeatother

\begin{document}

\begin{abstract}
It has been demonstrated that very simple attacks can fool highly-sophisticated neural network architectures.
In particular, so-called adversarial examples, constructed from perturbations of input data that are small or imperceptible to humans but lead to different predictions, may lead to an enormous risk in certain critical applications.
In light of this, there has been a great deal of work on developing adversarial training strategies to improve model robustness.
These training strategies are very expensive, in both human and computational time.
To complement these approaches, we propose a very simple and inexpensive strategy which can be used to ``retrofit'' a previously-trained network to improve its resilience to adversarial attacks. 
More concretely, we propose a new activation function---the JumpReLU---which, when used in place of a ReLU in an already-trained model, leads to a trade-off between predictive accuracy and robustness.
This trade-off is controlled by the jump size, a hyper-parameter which can be tuned during the validation stage. 
Our empirical results demonstrate that this increases model robustness, protecting against adversarial attacks with substantially increased levels of perturbations. 
This is accomplished simply by retrofitting existing networks with our JumpReLU activation function, without the need for retraining the model.
Additionally, we demonstrate that adversarially trained (robust) models can greatly benefit from retrofitting. 
\end{abstract}

\maketitle

\vspace{-0.5cm}

\section{Introduction}

\begin{figure}[h]
	\begin{center}
	\vspace{+0.5cm}
		\begin{overpic}[width=0.61\textwidth]{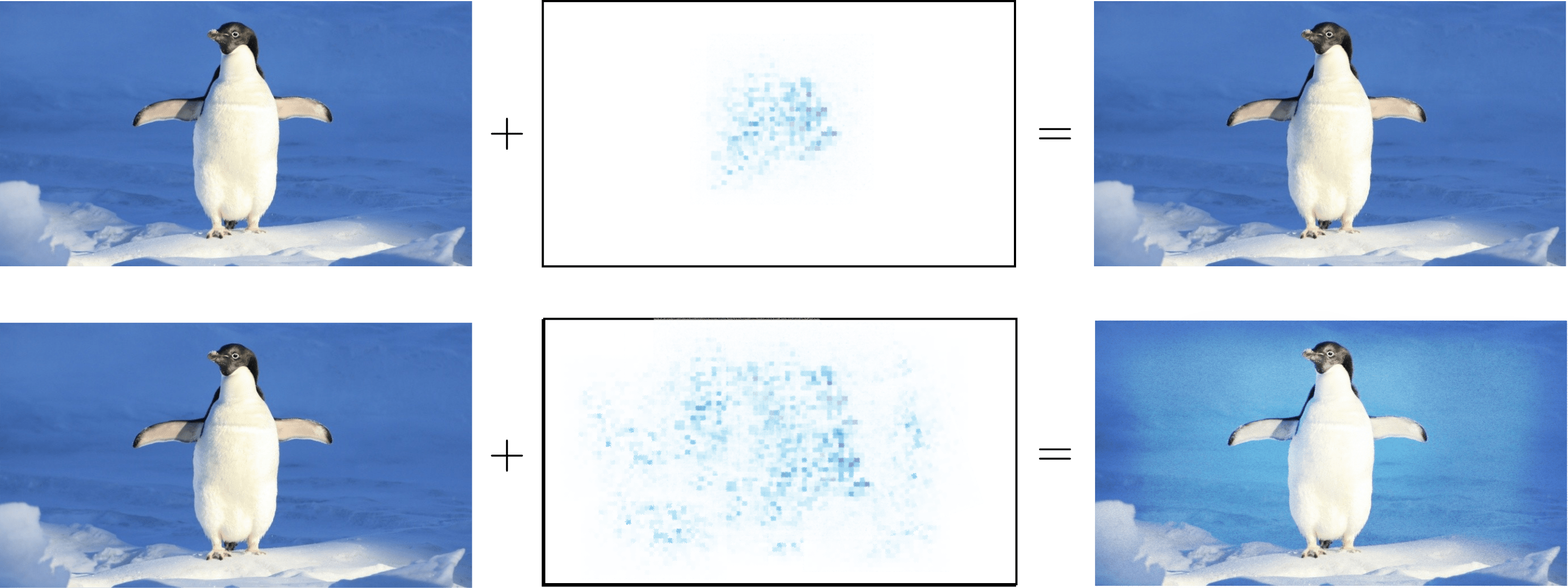} 
			%\put(-6,14){\rotatebox{90}{\small (a) Truth}}
			\put(5,40){{\footnotesize clean example}}
			\put(33.5,40){{\footnotesize adversarial perturbation}}
			\put(71,40){{\footnotesize adversarial example}}
		\end{overpic}	
		%\vspace{-0.3cm}
	\end{center}
	\caption{Adversarial examples are constructed by perturbing a clean example with a small amount of non-random noise in order to fool a classifier. Often, an imperceptible amount of noise is sufficient to fool a model (top row). The JumpReLU improves the robustness, i.e., a higher level of noise is required to fool the retrofitted model (bottom row). }
	\label{fig:illustration}
\end{figure}

%%From route optimization to determining medical diagnoses, neural networks have proven to be successful at solving complex, cognitive problems.
%
As machine learning methods become more integrated into a wide range of technologies, there is a greater demand for robustness, in addition to the usual efficiency and high-quality prediction, in machine learning algorithms.
Deep neural networks, in particular, are ubiquitous in many technologies that shape the modern world~\cite{lecun2015deep,goodfellow2016deep}, but it has been shown that even the most sophisticated network architectures can easily be perturbed and fooled by simple and imperceptible attacks.
For instance, single pixel changes which are undetectable to the human eye can fool neural networks into making erroneous predictions. 
These adversarial attacks can reveal important fragilities of modern neural networks~\cite{szegedy2013intriguing,Goodfellow2014ExplainingAH,liu2016delving}, and they can reveal flaws in network training and design which pose security risks~\cite{kurakin2016adversarial}. 
Partly due to this, evaluating and improving the robustness of neural networks is an active area of research. 
Due to the unpredictable and sometimes imperceptible nature of adversarial attacks, however, it can be difficult to test and evaluate network robustness comprehensively.
%
%%Here, we use the average minimum perturbation which is required to fool a model as a metric for robustness, illustrated in Figure~\ref{fig:illustration}.
See, \eg, Figure~\ref{fig:illustration}, which provides a visual illustration of how a relatively small adversarial perturbation can lead to incorrect classification.

Most work in this area focuses on training, \eg, developing adversarial training strategies to improve model robustness.  
These training strategies are very expensive, in both human and computational time. 
For example, a single training run can be expensive, and typically many training runs are needed, as the analyst ``fiddles with'' parameters and hyper-parameters.

\begin{figure*}[!t]
\begin{center}
\includegraphics[width=0.85\textwidth]{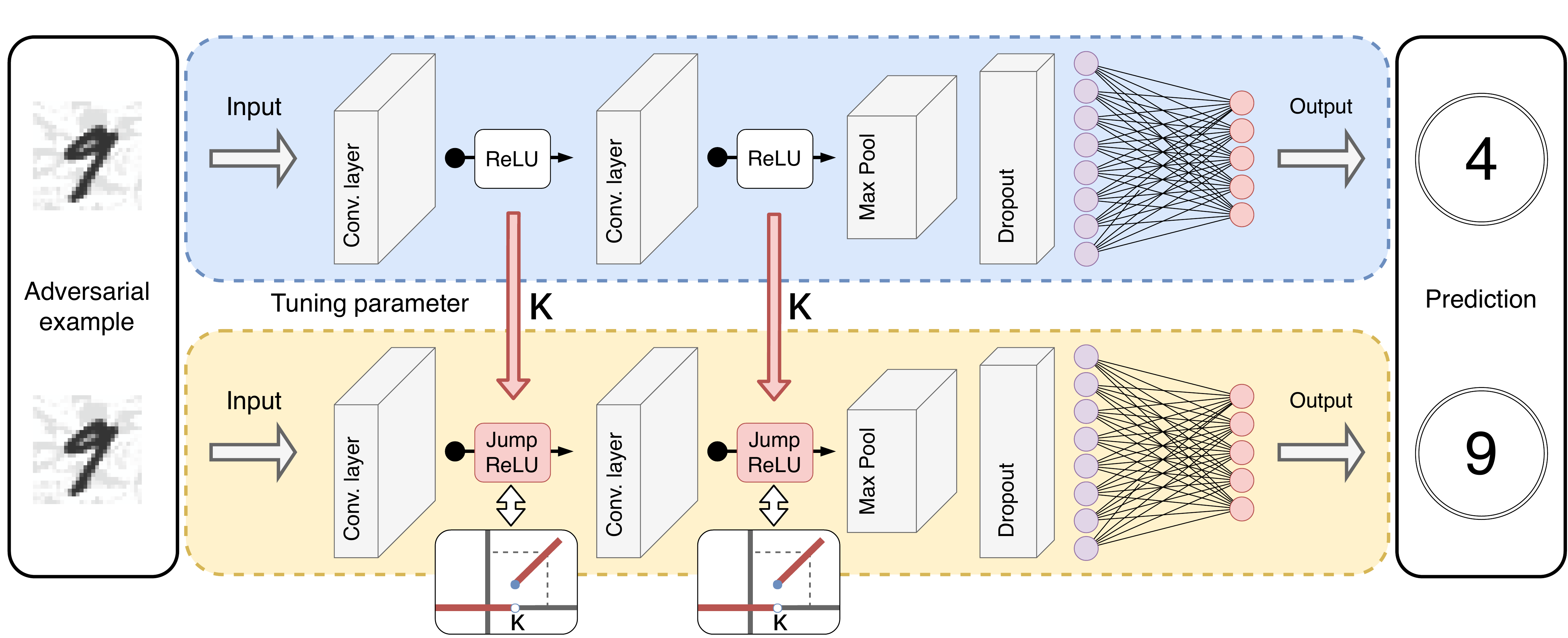}
\vspace{-0.4cm}
\end{center}
\caption{Simplified illustration of a neural network architecture using ReLU activation functions. JumpReLU can be activated by setting the jump value (threshold value) $\kappa$ larger than zero to increase the resilience to adversarial attacks.  (One could use different values of $\kappa$ for different layers, but we did not observe that to help.)}
\label{fig:train_vs_val}
\end{figure*}

Motivated by this observation, we propose a complementary approach to improve the robustness of the model to the risk of adversarial attacks. 
The rectified linear unit (ReLU) will be our focus here since it is the most widely-used and studied activation function in the context of adversarial attacks (but we expect that the same idea can be applied more generally).
For networks trained with ReLUs, our method will replace the ReLU with what we call a \emph{JumpReLU} function, a variant of the standard ReLU that has a jump discontinuity.
See Figure~\ref{fig:train_vs_val} for an illustration of the basic method. 
The jump discontinuity in the JumpReLU function has the potential to dampen the effect of adversarial perturbations, and we will ``retrofit'' existing ReLU-based networks by replacing the ReLU with the JumpReLU, as a defense strategy, to reduce the risk of adversarial attacks.
The magnitude of the jump is a parameter which controls the trade-off between predictive accuracy and robustness, and it can be chosen in the validation stage, \ie, without the need to retrain the network.

In more detail, our contributions are the following:
 \begin{itemize}
 \setlength{\itemsep}{1pt}
 \setlength{\parskip}{0pt}
 \setlength{\parsep}{0pt}

 \item We introduce and propose the \emph{JumpReLU activation function}, a novel rectified linear unit with a small jump discontinuity, in order to improve the robustness of trained neural networks.
 
 \item We show that the JumpReLU activation function can be used to ``retrofit'' already deployed, \ie, pre-trained, neural networks---\emph{without the need to perform an expensive retraining of the original network}. 
 Our empirical results show that using the JumpReLU in this way leads to networks that are resilient to substantially increased levels of perturbations, when defending classic convolutional networks and modern residual networks.
 We also show that JumpReLU can be used to enhance adversarially trained (robust) models. 
 %
 %We also demonstrate the robustly trained models can benefit from being retrofitted. 
 %
  
 \item  We show that the popular Deep Fool method requires increased noise levels by a factor of about $3$--$7$ to achieve nearly $100$ percent fooling rates for the retrofitted model on CIFAR10. We show that these increased noise levels are indeed critical, \ie, the detection rate of adversarial examples is substantially increased when using an additional add-on detector.

 \item The magnitude of the jump is an additional hyper-parameter in the JumpReLU activation function that provides a trade-off between predictive accuracy and robustness.
 This single parameter can be efficiently tuned during the validation stage, \ie, without the need for network retraining.

 %defense distillation~\cite{hinton2015distilling, papernot2016distillation}. % as well as transferability properties~\cite{papernot2016transferability}. 
 \end{itemize}

In summary, the JumpReLU activation functions improves the model robustness to adversarial perturbations, while attaining a ``good'' accuracy for clean examples. Further, the impact on the architecture is minimal and does not effect the inference time of the network.  

% \begin{figure}[!b]
% \begin{center}
% \includegraphics[width=0.4\textwidth]{figs/overview.pdf}
% \vspace{-0.6cm}
% \end{center}
% \caption{Good caption.}
% \label{fig:overview}
% \end{figure}

\section{Related work}

Adversarial examples are an emerging threat for many machine learning tasks.
%%, designed with the intention to produce a beneficial outcome for the attacker.
%
%
Szegedy \etal~\cite{szegedy2013intriguing} discovered that neural networks are particularly susceptible to such adversarial examples. 
This can lead to problems in safety- and security-critical applications such as medical imaging, surveillance, autonomous driving, and voice command recognition. Due to its importance, adversarial learning has become an intense area of research, posing a cat-and-mouse game between attackers and defenders. 

Indeed, there is currently a lack of theory to explain why deep learning is so sensitive to this form of attack.  
Early work hypothesized that the highly non-linear characteristics of neural networks and the tendency toward almost perfect interpolation of the training data are the reasons for this phenomena.  
Tanay and Griffin~\cite{tanay2016boundary} argued that the adversarial strength is related to the level of regularization and that the effect of adversarial examples can be mitigated by using a proper level of regularization.
In contrast, Goodfellow \etal~\cite{Goodfellow2014ExplainingAH} impressively demonstrated that the linear structure in deep networks with respect to their inputs is sufficient to craft adversarial examples.   
%

%Akhtar and Mian~\cite{akhtar2018threat} provide a survey of the threat of adversarial attacks on deep learning in computer vision. 

Let's assume that $x$ denotes an input such as an image. The problem of crafting an adversarial example $\tilde{x}$ requires finding an additive perturbation $\Delta x$, so that $\tilde{x}$ which is constructed as
\begin{equation}
    \tilde{x} \, = \, x  \, + \, \Delta x,
\end{equation}
fools a specific model $F$ under attack. 
The minimal perturbation with respect to a $p$-norm $\|\cdot\|_p$ can be obtained by using an optimization based strategy which aims to minimize
\begin{equation}
\Delta x \, :=  \,\argmin_{\Delta \hat{x}} \|\Delta \hat{x} \|_p \quad \text{s.t.} \quad F(x + \Delta \hat{x}) \ne F(x),
\end{equation}
so that the example $x$ is misclassified.

Note, the perturbation used to construct  adversarial examples needs to be small enough to be unnoticeable for humans, or add-on detection algorithms. Intuitively, the average minimum perturbation which is required to fool a given model yields a plausible metric to characterize the robustness of a model~\cite{papernot2016distillation}. Hence, we can quantify the robustness for a trained model $F$ as
\begin{equation}
\rho_{F} \, := \, \mathop{\mathbb{E}_{(X,Y) \sim \mathcal{D}}} \, \left[ \frac{\|\Delta X\|_p}{\|X\|_p} \right],
\end{equation}
where the input-target-pairs $(X,Y)$ are drawn from distribution $\mathcal{D}$, and $\Delta X$ is the minimal perturbation that is needed to fool the model $F$.

\subsection{Attack strategies}

There are broadly two types of attacks: targeted and non-targeted attacks. 
Targeted attacks aim to craft adversarial examples which fool a model to predict a specific class label. 
Non-targeted attacks have a weaker objective, \ie, simply to classify an adversarial example incorrectly.

Independent of the type, attack strategies can be categorized broadly into two families of threat models.
\textit{Black-box attacks} aim to craft adversarial examples without any prior knowledge about the target model~\cite{su2017one,sarkar2017upset,cisse2017houdini,dong2017boosting}. 
\textit{White-box attacks}, in contrast, require comprehensive prior knowledge about the target model. 
There are several popular white-box attacks for computer vision applications~\cite{szegedy2013intriguing,Goodfellow2014ExplainingAH,liu2016delving,moosavi2016deepfool,kurakin2016adversarial,moosavi2017universal,poursaeed2017generative}. 
A slightly weaker form are \textit{gray-box} attacks, which take advantage of partial knowledge about the target model.

The following (non-targeted) attack methods are particularly relevant for our results.
\begin{itemize}
\setlength{\itemsep}{1pt}
\setlength{\parskip}{0pt}
\setlength{\parsep}{0pt}

\item
First, the Fast Gradient Sign Method (FGSM)~\cite{Goodfellow2014ExplainingAH}, which crafts adversarial perturbations $\Delta x$ by using the sign of the gradient of the loss function $\mathcal{L}$ with respect to the clean input image $x$. Let's assume that the true label of $x$ is $y$. Then, the adversarial example $\tilde{x}$ is constructed as
\begin{equation}
    \tilde{x} = x + \epsilon \cdot \sign(\nabla_x \mathcal{L}(F(x), y)),
\end{equation}
where $\epsilon$ controls the magnitude of the perturbation. Here, the operator $\sign$ is an element-wise function, extracting the sign of a real number. 

Relatedly, the iterative variant IFGSM~\cite{kurakin2016adversarial} constructs adversarial examples using $1,...,k$ iterations
\begin{equation}
\tilde{x}_k = \clip_{x} \left[ \tilde{x}_{k-1} + \epsilon  \cdot  \sign(\nabla  \mathcal{L}( F(\tilde{x}_{k-1}), y)) \right],
\end{equation}
where $\clip_{x}$ is an element-wise clipping function.  This approach is essentially a projected gradient decent (PGD) method used to craft adversarial examples~\cite{madry2017towards}.
\item
Second, the Deep Fool (DF) method, which is another iterative method for constructing adversarial examples~\cite{moosavi2016deepfool}. 
The DF method first approximates the model under consideration as a linear decision boundary, and then seeks the smallest perturbation needed to push an input image over that boundary. 
DF can minimize the loss function using either the $L_\infty$ or  $L_2$ norm.
\item
Third, the recently introduced trust region (TR) based attack method~\cite{yao2018trust}. 
In~\cite{yao2018trust}, the authors show that this TR method performs similarly to the Carlini and Wagner (CW)~\cite{carlini2017towards} attack method, but is more efficient in terms of the computational resources required to construct the adversarial examples. 
%This attack method is a new optimization-based technique for constructing adversarial examples. 
%Empirical results demonstrate that the CW method is among the most effective attack strategies, but it is computationally demanding to construct adversarial examples using this method. 
%Thus, we consider the CW method only for constructing a few adversarial examples.   
%
\end{itemize}

\subsection{Defense strategies}
Small perturbations are often imperceptible for both humans and the predictive models, making the design of counterattacks a non-trivial task. 
%
%In practice, c
Commonly used techniques for preventing overfitting (\eg, including weight decay and dropout layers and then retraining) do not robustify the model against adversarial examples.
Akhtar and Mian~\cite{akhtar2018threat} segment modern defense strategies into three categories.

The first category includes strategies which rely on specialized add-on (external) models which are used to defend the actual network~\cite{akhtar2017defense,lee2017generative,shen2017ape,xu2017feature,meng2017magnet}. 

The second category includes defense strategies which modify the network architecture in order to increase the robustness~\cite{gu2014towards,ross2017improving,papernot2016distillation,nayebi2017biologically,guo2017countering}.
Closely related to our work, Zantedeschi \etal~\cite{zantedeschi2017efficient} recently proposed a bounded ReLU activation function as an efficient defense against adversarial attacks.
Their motivation is to dampen large signals to prevent accumulation of the adversarial perturbation over layers as a signal propagates forward, using the function.
   
The third category aims to modify the input data for the training and validation stage in order to improve the robustness of the model~\cite{miyato2016adversarial,zheng2016improving,guo2017countering,bhagoji2018enhancing,luo2015foveation,liao2017defense,prakash2018deflecting}.    
    
A drawback of most state-of-the-art defense strategies is that they involve modifying the network architecture.
Such strategies require that the new network is re-trained or that new specialized models are trained from scratch. 
This retraining is expensive in both human and computation time.
Further, specialized external models can require considerable effort to be deployed and often increase the need of computational resources and inference time. 

\section{Jump rectified linear unit (JumpReLU)}

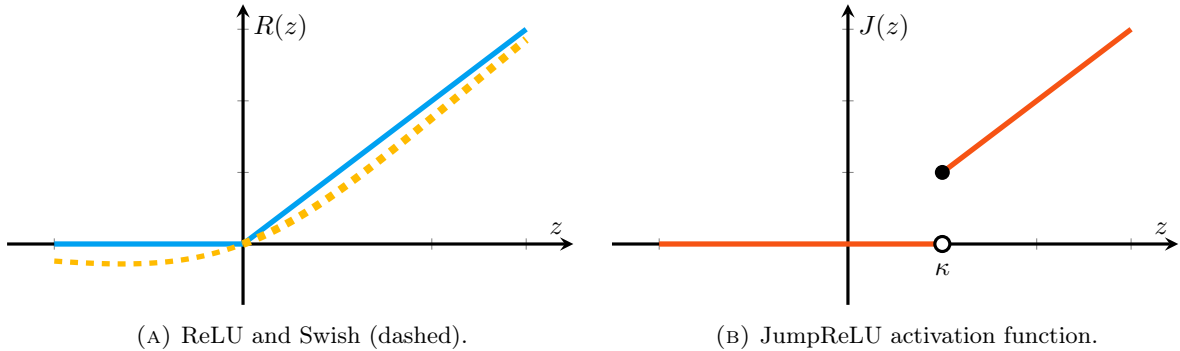
\begin{figure}[!t]
	\centering
	
	\begin{subfigure}[t]{0.48\textwidth}
		\begin{tikzpicture}
		\begin{axis}[
		very thick,
		axis lines = middle,
		enlargelimits = true,
		width=.95\textwidth, height=4cm,
		xmin=-2, xmax=3, ymin=-0.5, ymax=3,
		no markers,
		samples=50,
		axis lines*=left, 
		axis lines*=middle, 
		scale only axis,
		xtick={-2,  0,  2, 3},
		ytick={ 0, 1, 2, 3},
		xticklabels={},
		yticklabels={},		
		xlabel={$z$},
		ylabel={$R(z)$}	
		] 
		%\addplot[red,domain=-2:-1,densely dashed]{-x-.5};
		\addplot[line width=2.0pt, mblue, domain=-2:0]{0};
		\addplot[line width=2.0pt, mblue, domain=0:+3]{x};
		
		\addplot[dashed, line  width=2.0pt, morange, domain=-2:0]{x /(1+exp(-1*1*x)};
		\addplot[dashed, line width=3.0pt, morange, domain=0:+3]{x /(1+exp(-1*1*x)};		
		
		\end{axis}
		%\draw[gray, dashed] (5.6,1)--(5.6,2.7);
		%\draw[gray, dashed] (5.6,2.7)--(3.1,2.7);
		\end{tikzpicture}
		\caption{ReLU and Swish (dashed).}
		\label{fig:ReLU}
	\end{subfigure}
	~
	\begin{subfigure}[t]{0.48\textwidth}
		%\vspace{+.40in}
		\begin{tikzpicture}
		\begin{axis}[
		very thick,
		axis lines = middle,
		enlargelimits = true,
		width=.95\textwidth, height=4cm,
		xmin=-2, xmax=3, ymin=-0.5, ymax=3,
		no markers,
		samples=50,
		axis lines*=left, 
		axis lines*=middle, 
		scale only axis,
		xtick={-2,  0,  2, 3},
		ytick={0, 1, 2, 3},
		xticklabels={},
		yticklabels={},			
		xlabel={$z$},
		ylabel={$J(z)$}	
		] 
		\addplot[line width=2.0pt, mred, domain = -2:1]{0};
		\addplot[line width=2.0pt, mred, domain = 1:+3]{x};
		
		%\addplot[dashed, line  width=2.0pt, morange, domain=-2:1]{0};
		%\addplot[dashed, line width=3.0pt, morange, domain=1:+3]{x /(1+exp(-1*1*x)};		

		\node[label={180:{}},circle,fill,inner sep=2pt] at (axis cs:1,1) {};
		\node[label={below:{$\kappa$}},circle,draw=black,fill=white,inner sep=2pt] at (axis cs:1,0) {};  
		
		%\node[label={180:{}},circle,fill,inner sep=2pt] at (axis cs:1,0.7) {};

		\end{axis}
		
		%\draw[gray, dashed] (4.35, 1.0)--(4.35,1.75);
		
		%\draw[gray, dashed] (3.1,1.75)--(4.35,1.75);        
		
		%\draw[gray, dashed] (5.6,1)--(5.6,2.7);
		%\draw[gray, dashed] (5.6,2.7)--(3.1,2.7);        
		%\draw[gray, dashed] (5.6,2.7)--(3.1,2.7); 
		\end{tikzpicture}
		\caption{JumpReLU activation function.}
		\label{fig:JumpReLU}
	\end{subfigure}

	\caption{The rectified linear unit is the most widely studied activation function in context of adversarial attacks, illustrated in (a). In addition its smooth approximation (Swish) is shown, with $\beta=0.5$. The JumpReLU activation function (b) introduces robustness and an additional amount of sparsity, controlled via the jump value (threshold value) $\kappa$. In other words, JumpReLU suppresses small positive signals. }
\end{figure}

The rectified linear unit (ReLU) and its variants have arguably emerged as the most popular activation functions for applications in the realm of computer vision. 
The ReLU activation function has beneficial numerical properties, and also has sparsity promoting properties~\cite{glorot2011deep}.
Indeed, sparsity is a widely used concept in statistics and signal processing~\cite{hastie2015statistical}.
%%, as well as in computational neuroscience~\cite{olshausen1997sparse}. 
%
For a given input $x$ and an arbitrary function $f:\mathcal{R}\xrightarrow{}\mathcal{R}$, the ReLU function can be defined as the positive part of the filter output $z = f(x)$ as
\begin{equation}
    R(z) := \max(z,0),
\end{equation}
illustrated in Figure~\ref{fig:ReLU}. The ReLU function is also known as the ramp function which has several other interesting definitions. For instance, we can define the ReLU function as
\begin{equation}
    R(z) := z H(z),
\end{equation}
where $H$ is the discrete Heaviside unit step function
\begin{equation}
    H(z) := \begin{cases}
    0 \quad \text{if} \,\, z \leq 0,\\
    1 \quad \text{if} \,\, z > 0.
\end{cases}
\end{equation}
Alternatively, the logistic function can be used for smooth approximation of the Heaviside step function
\begin{equation}
    H(z) :\approx \frac{1}{1 + \exp{(-2\beta z)}}.
\end{equation}
Intriguingly, this smooth approximation resembles the Swish activation function~\cite{ramachandran2018searching}, which is defined as
\begin{equation}
    S(z) := z \frac{1}{1 + \exp{(-2\beta z)}}.
\end{equation}

The ReLU activation function works extremely well in practice.
However, 
%%from a biophysical perspective, 
a fixed threshold value $0$ seems 
%to be 
arbitrary. 
%% %
%% Looking to biophysical systems can inspire more biologically plausible as well as more robust neural networks and activation functions~\cite{glorot2011deep,nayebi2017biologically}. 
%% %%%
%% In particular, the brain utilizes nonlinear activation functions to filter out noise sources in the neural response. Biophysical properties inherent to each neuron determines a threshold for eliciting an action potential~\cite{amit_1989}. 
%% %This allows the neuron to collect and transmit relevant information. 
%% %%%
%% %
%% %Neurons gate their electrical signaling by setting response-thresholds based on the intrinsic properties of the cell and the dynamics of the input signal~\cite{amit_1989}. 
%% %
%% Indeed, thresholding filters out noise sources in the neural response. This allows the neuron to collect and transmit relevant information. 
%% %
Thus, it seems reasonable to crop activation functions 
%in artificial neural networks 
so that they turn on only for inputs greater than or equal to the jump value $\kappa$. 
%As a consequence, 
In this case, 
sub-threshold signals are suppressed, while significant signals are allowed to pass.
%
%In other words, JumpReLU filters signals which have a negative sign as well as signals of small positive magnitude.

We introduce the JumpReLU function which suppresses signals of small magnitude and negative sign
\begin{equation}
    J(z) := z H(z-\kappa) = 
    \begin{cases}
    0 \quad  \text{if} \,\, z \leq \kappa\\
    z \quad  \text{if} \,\, z > \kappa,
\end{cases}
\end{equation}
illustrated in Figure~\ref{fig:JumpReLU}. 
This activation function introduces a jump discontinuity, yielding piece-wise continuous functions. 
While this idea can likely be transferred to other activation functions, we restrict our focus to the family of discrete ReLU activation functions.

Glorot \etal~\cite{glorot2011deep} note that too much sparsity can negatively affect the predictive accuracy. 
Indeed, this might be an issue during the training stage, however, a fine-tuned jump value $\kappa$ can improve the robustness of the model during the validation stage by introducing an extra amount of sparsity.
The tuning parameter $\kappa$ can be used to control the trade-off between predictive accuracy and robustness of the model. 
%
%Sub-threshold signals are suppressed are more prone to perturbations. 
%
%The positive effect of the JumpReLU activation function is pronounced for increasing amounts of external noise or perturbations of the stimuli or inputs.
%
%
Importantly, JumpReLU can be used to retrofit previously trained networks in order to mitigate the risk to be fooled by adversarial examples.   

Note that the jump value $\kappa$ can be tuned cheaply during the validation stage once the network is trained, \emph{i.e.}, without the need for expensive retraining of the model.

%Zantedeschi \etal~\cite{zantedeschi2017efficient} recently proposed a bounded ReLU activation as an efficient defense against adversarial attacks.
%%
%Their motivation is to dampen large signals to prevent accumulation of the adversarial perturbation over layers as a signal propagates forward, using the function
%%
%\begin{equation}
%    B(z) :=  
%    \begin{cases}
%    0 \quad  \text{if} \,\, z < 0\\
%    z \quad   \text{if} \,\, 0 \leq z < \theta\\
%    \theta \quad  \text{if} \,\, z \geq \theta.
%\end{cases}
%\end{equation}
%%
%The bounded ReLU activation function is illustrated in Figure~\ref{fig:BoundedReLU}.
%%
%While~\cite{zantedeschi2017efficient} proposed the bounded ReLU activation function for training, we are interested if this activation can also be used as a retrofit strategy. 

% 16.5 / 29.6 30.1 46.9

\section{Experiments}\label{sec:tb_result}

We first outline the setup which we use to evaluate the performance of the proposed JumpReLU activation function.\footnote{Reserach code is available here:  \url{https://github.com/erichson/JumpReLU}.} We restrict our evaluation to MNIST and CIFAR10, since these are the two standard datasets which are most widely used in the literature to study adversarial attacks. 
%In summary, we use the following setup:
%

\begin{itemize}
  \setlength{\itemsep}{1pt}
  \setlength{\parskip}{0pt}
  \setlength{\parsep}{0pt}

    \item The \textbf{MNIST} dataset~\cite{lecun-gadientbasedlearning} provides $28\times28$ gray-scale image patches for 10 classes (digits), comprising $60,000$ instances for training, and $10,000$ examples for validation. 
    For our experiments we use a LeNet5 architecture with an additional dropout layer, which we denote as LeNetLike.%, which achieves nearly state-of-the-art accuracy. 
    
    \item The \textbf{CIFAR10} dataset~\cite{krizhevsky2009learning} provides $32\times32$ RGB image patches for 10 classes, comprising $50,000$ instances for training, and $10,000$ examples for validation. 
    For our CIFAR10 experiments we use a simple AlexLike architecture proposed by~\cite{carlini2017towards}; a wide residual network (WideResNet) architecture~\cite{zagoruyko2016wide} of depth 30 and with width factor $4$; and a MobileNetV2 architecture which is using inverted residuals and linear bottlenecks~\cite{sandler2018mobilenetv2}. %a ResNet architecture of depth 20~\cite{he2016deep};

    %\item \textbf{ImageNet} (ILSVRC 2012)~\cite{deng2009imagenet} is the most challenging data set that we consider. ImageNet comprises $1,000$ classes and a total of $1.2$ million training and $50,000$ test images. Here, we use the VGG16 architecture~\cite{simonyan2014very} as a classifier. Note, that the network is trained using cropped images of dimension $224 \times 224$.
\end{itemize}

We aim to match the experimental setup for creating adversarial examples as closely as possible to prior work. 
Thus, we follow the setup proposed by Madry \etal~\cite{madry2017towards} and Buckman \etal~\cite{buckman2018thermometer}. 
More concretely, we use for all MNIST experiments $\epsilon = 0.01$ and $40$ steps for iterative attacks; for experiments on CIFAR10 we use the same $\epsilon$, and $7$ steps for PGD and Deep Fool attacks. For the trust region attack method we use $1000$ steps. Note, these values are chosen by following the assumption that an attacker aims to construct adversarial examples which have imperceptible perturbations. Further, we assume that the attacker has only a limited budget of computational resources at disposal. 
%Of course, it would be a trivial task to fool any model using adversarial examples which have an arbitrary large perturbation pattern.  

We also evaluate the effectiveness of the JumpReLU for adversarially trained networks. Training the model with adversarial examples drastically increases the robustness with respect to the specific attack method used to generate the adversarial training examples~\cite{Goodfellow2014ExplainingAH}. Here, we use the FGSM method to craft examples for adversarial training with $\epsilon=0.3$ for MNIST, and $\epsilon=0.03$ for CIFAR10.  Unlike Madry \etal~\cite{madry2017towards}, we perform robust training with mixed batches composed of both clean and adversarial examples. This leads to an improved accuracy on clean examples, while being slightly less robust. 
%(Supplementary materials provide additional experiments for models trained purely on adversarial inputs.) 
The specific ratio of the numbers of clean to adversarial examples can be seen as a tuning parameter, which may depend on the application.

\subsection{Results}

In the following, we compare the performance of JumpReLU to the standard ReLU activation function for both gray-box and white-box attack scenarios. For each scenario, we consider three different iterative attack methods: the Projected Gradient Descent (PGD) method, the Deep Fool (DF) method using both the $L_2$ (denoted as DF$_2$) and $L_\infty$ norm (denoted as DF$_\infty$) as well as the Trust Region (TR) attack method.\footnote{We use the TR method as a surrogate for the the more popular Carlini and Wagner (CW)~\cite{carlini2017towards} attack method. This is because the CW method requires enormous amounts of computational resources to construct adversarial examples. For instance, it takes about one hour to construct $300$ adversarial examples for CIFAR10 using the CW method, despite using a state-of-the-art GPU and the implementation provided by~\cite{rauber2017foolbox}. Yao \etal.~\cite{yao2018trust} show that the TR method requires similar average and worst case perturbation magnitudes as the CW method does in order to attack a specific network.}

\subsubsection{Gray-box attack scenario}
We start our evaluation by considering the gray-box attack scenario. In this ``vanilla'' flavored setting, we assume that the adversary has only partial knowledge about the model. Here, the adversary has full access to the ReLU network to construct adversarial examples, but it has no information about the JumpReLU setting during inference time. In other words, the ReLU network is used as a source network to craft adversarial examples which are then used to attack the JumpReLU network. 
We present results for models trained on clean data only (base) and adversarially trained models (robust). 
Table~\ref{tab:graybox} shows a summary of results for MNIST and CIFAR10 using different network architectures. 
The positive benefits of JumpReLU are pronounced, while the loss of accuracy on clean examples is moderate. 

First, Tab.~\ref{tab:graybox_mnist} shows for MNIST that the retrofitted models have a substantially increased resilience to gray-box attacks. 
Especially the adversarial examples, which are crafted using the PGD method, turn out to be ineffective for fooling both the retrofitted base and robust (highlighted in gray) models. Further, we can see that the JumpReLU increases the resilience to the DF and TR attack methods.

Next, Tables~\ref{tab:graybox_alex},~\ref{tab:graybox_wide}, and~\ref{tab:graybox_mobile} show results for CIFAR10. 
Clearly, the more complex residual networks (Tab.~\ref{tab:graybox_wide} and Tab.~\ref{tab:graybox_mobile}) appear to be more vulnerable than the simpler AlexLike network (Tab.~\ref{tab:graybox_alex}).
The JumpReLU is able to prevent the PGD gray-box attack on the AlexLike network, whereas the stand-alone JumpReLU is insufficient to defend the base WideResNet and MobileNetV2. %However, the JumpReLU increases the resilience with respect to the DF and TR attack methods. 
Still, JumpReLU is able to substantially increase the robustness with respect to the Deep Fool and Trust Region attacks. 

Surprisingly, the JumpReLU is able to substantially improve the resilience of robustly trained models. In case of the PGD gray-box attack, the retrofitted model improves the accuracy from $60.43\%$ to $70.25\%$ for the WideResNet (Tab.~\ref{tab:graybox_wide}) and from $53.98\%$ to $66.37\%$ for the MobileNetV2 (Tab.~\ref{tab:graybox_mobile}). 
Indeed, this demonstrates the flexibility of our approach and shows that retrofitting is not limited to weak models only. 

While we see that the adversarially trained models are more robust with respect to the specific attack method used for training, it can also be seen that such models provide no significant protection for other attack methods. 
In contrast, our defense strategy  based on the JumpReLU is agnostic to specific attack methods, \emph{i.e.}, we improve the robustness with respect to all attacks considered here.  Note we could further increase the jump value for the robust models, in order to increase the robustness to the Deep Fool and TR attack method. However, this comes with the price of sacrificing slightly more accuracy on clean data.

Appendix~\ref{app:second_graybox} provides additional results for the gray-box attack scenario, showing that the crafted adversarial examples are ``unidirectional,'' in the sense that adversarial examples crafted by using source models which have a low jump value can be used to attack models which have a higher jump value, but not vice versa.

\begin{table}[!t]
	\caption{Summary of results for black-box attacks. The numbers indicate the accuracy, \ie, the percentage of correctly classified instances (higher numbers indicate better robustness). Here, the ReLU network (indicated by a `*') is used as the source model to generate adversarial examples.\label{tab:graybox} }
	\setlength\tabcolsep{3.2pt}
	\begin{subtable}{1\textwidth}
	\centering
	\scalebox{0.99}{
		\begin{tabular}{lcccccccccc} \toprule
			Model            	& Accuracy & PGD  & DF$_\infty$    &DF$_2$   &TR\\
			\midrule
			ReLU  (Base)*   & \textbf{99.55\%}  & 66.69\%    & 0.0\%             & 0.0\%            &0.0\%   \\ 
			JumpReLU (Base)     		& 99.53\%   		& 91.65\%    & \textbf{81.39\%}  & \textbf{58.93\%} &\textbf{58.90}\\
				
			\midrule 
			\rowcolor{graytable}
			ReLU  (Robust)* & 99.50\%  & 91.39\%          & 0.0\%   & 0.0\%    &0.0\%\\
			\rowcolor{graytable}
			JumpReLU (Robust)       & 99.47\%  & \textbf{97.07\%} & 70.84\% & 45.17\%  &53.24\%\\
		
			\bottomrule 
	\end{tabular}}
	%\captionsetup{labelformat=empty}
	\caption{Results for LeNetLike network (MNIST); $\kappa = 1.0$.}
    \label{tab:graybox_mnist}
    \end{subtable}

	\begin{subtable}{1\textwidth}
	\centering
	%\vspace{-0.1cm}
	\scalebox{0.99}{
		\begin{tabular}{lcccccccccc} \toprule
			Model            	& Accuracy   & PGD & DF$_\infty$ &DF$_2$ &TR\\
			\midrule
	ReLU  (Base)* & \textbf{89.46\%}  & 6.38\%  & 0.0\%            & 0.0\%            &0.0\%\\ 
	JumpReLU (Base)         & 87.52\%           & 45.75\% & \textbf{61.82\%} & \textbf{60.55\%} &\textbf{53.08\%}\\
			
			\midrule 
			\rowcolor{graytable}
	ReLU  (Robust)* & 87.93\%   & 51.88\%          & 0.0\%   & 0.0\%  &0.0\%  \\ 
			\rowcolor{graytable}
	JumpReLU (Robust)       & 86.19\%   & \textbf{67.65\%} & 52.28\% & 46.9\% &51.52\%  \\
			
			\bottomrule 
	\end{tabular}}
	%\captionsetup{labelformat=empty}
	\caption{Results for AlexLike network (CIFAR10); $\kappa = 0.4$.}
    \label{tab:graybox_alex}
	\end{subtable}

	\begin{subtable}{1\textwidth}
	\centering
	%\vspace{-0.1cm}
	\scalebox{0.99}{
	\begin{tabular}{lcccccccccc} \toprule
		Model            & Accuracy  & PGD    & DF$_\infty$   &DF$_2$  &TR\\
		\midrule
		ReLU  (Base)*  & \textbf{94.31\%} & 0.0\%  & 0.0\%            & 0.0\%                &0.0\%\\
		JumpReLU  (Base)        & 92.58\%    		 & 0.39\% & \textbf{37.33\%} & \textbf{40.21\%} & \textbf{45.90\%}\\
		
		\midrule 
		\rowcolor{graytable}
		ReLU  (Robust)*  & 93.72\% & 60.43\%         	 & 0.0\%   & 0.0\%    &0.0\%\\    
		\rowcolor{graytable}
		JumpReLU (Robust)       & 93.01\% & \textbf{70.25\%} & 28.62\% & 26.11\%  &35.33\% \\
		\bottomrule 
	\end{tabular}}
	%\captionsetup{labelformat=empty}
	\caption{Results for WideResNet (CIFAR10); $\kappa = 0.07$.}
    \label{tab:graybox_wide}
	\end{subtable}

	\begin{subtable}{1\textwidth}
	\centering
	%\vspace{-0.1cm}
	\scalebox{0.99}{
	\begin{tabular}{lcccccccccc} \toprule
		Model        & Accuracy  & PGD    & DF$_\infty$ &DF$_2$\\
		\midrule
		ReLU  (Base)*     & \textbf{92.07\%}  & 0.0\%  & 0.0\%            & 0.0\%            &0.0\%\\
		JumpReLU (Base)        & 90.43\%    		 & 0.54\% & \textbf{40.69\%} & \textbf{41.61\%} &\textbf{43.18\%}\\

		\midrule 
		\rowcolor{graytable}
		ReLU  (Robust)*  & 91.69\%  & 53.98\%         	 & 0.0\%    & 0.0\%    &0.0\%\\ 
		\rowcolor{graytable}
		JumpReLU (Robust)        & 90.12\%  & \textbf{66.37\%} & 37.31\% & 35.45\%  &40.4\% \\
		\bottomrule 
	\end{tabular}}
	%\captionsetup{labelformat=empty}
	\caption{Results for MobileNetV2 (CIFAR10); $\kappa = 0.06$.}
    \label{tab:graybox_mobile}
	\end{subtable}

\vspace{-0.6cm}		
\end{table}

\begin{table}[!b]
%\vspace{-0.4cm}	
	\caption{Summary of results for white-box attacks. The numbers indicate the accuracy, \ie, the percentage of correctly classified instances (higher numbers indicate better robustness). The Deep Fool method is able to fool all instances using only $7$ iterations, hence we show here the average minimum perturbations in parentheses. The best performance in each category is highlighted in bold letters.\label{tab:whitebox}}
	\setlength\tabcolsep{3.2pt}
	
	\begin{subtable}{1\textwidth}
	\centering
	\scalebox{0.99}{
		\begin{tabular}{lcccccccccc} \toprule
			Model              & Accuracy    & PGD  & DF$_\infty$  &DF$_2$  &TR\\
			\midrule
			ReLU  (Base)   & \textbf{99.55\%} & 66.69\% & (17.9\%) & (21.8\% ) & (18.9\%) \\            
			JumpReLU (Base)     	   & 99.53\%          & 83.21\% & (34.1\%) & (44.9\% ) & (25.0\%)\\
			
			\midrule 
			\rowcolor{graytable}
			ReLU  (Robust)    & 99.50\%  & 91.39\%          & (28.4\% )          & (31.4\% )          &(24.7\%)\\ 
			\rowcolor{graytable}
			JumpReLU (Robust)     	  & 99.47\%  & \textbf{94.36\%} & (\textbf{46.6\%}) & (\textbf{53.3\%}) &(\textbf{32.8\%}) \\
			\rowcolor{graytable}
			\midrule 			
			\rowcolor{graytable}
			Madry~\cite{madry2017towards}			 & 98.80\%           	& 93.20\%      & -        &-        		& -  \\
			\rowcolor{graytable}
			Vanilla~\cite{buckman2018thermometer}  & 99.03\%           	& 91.36\%             & -       &-        		& -  \\				
			\rowcolor{graytable}
			One-hot~\cite{buckman2018thermometer}  & 99.01\%           	& 93.77\%             & -         &-      		& -  \\
			\rowcolor{graytable}			
			Thermo~\cite{buckman2018thermometer} & 99.23\%           	& 93.70\%            & -            &-    		& -  \\	 		
			
			\bottomrule 
	\end{tabular}}
	%\captionsetup{labelformat=empty}
	\caption{Results for LeNetLike network (MNIST); $\kappa = 1.0$.}
    \label{tab:whitebox_mnist}
	\end{subtable}	
	
	\begin{subtable}{1\textwidth}
	\centering
	\scalebox{0.99}{
		\begin{tabular}{lcccccccccc} \toprule
			Model            	& Accuracy     & PGD   & DF$_\infty$ &DF$_2$ &TR\\
			\midrule
			ReLU  (Base)    & \textbf{89.46\%}	& 6.38\%  & (1.2\%)	 & (1.5\%)  &(1.3\%) \\                
			JumpReLU (Base)           & 87.52\%           & 18.56\% & (9.80\%) & (10.6\%) &(1.7\%) \\
			
			\midrule 
			\rowcolor{graytable}
	ReLU  (Robust) & 87.93\% & 51.88\%          & (3.6\%)           & (4.2\%)           &(3.6\%)\\
			\rowcolor{graytable}
	JumpReLU (Robust)      & 86.19\% & \textbf{56.70\%} & (\textbf{13.2\%}) & (\textbf{14.1\%}) &(\textbf{4.3\%})  \\
			\bottomrule 
	\end{tabular}}
	\caption{Results for AlexLike network (CIFAR10); $\kappa = 0.4$.}
    \label{tab:whitebox_alex}
	\end{subtable}
	
	\begin{subtable}{1\textwidth}
	\centering
	\scalebox{0.99}{
		\begin{tabular}{lcccccccccc} \toprule
			Model            & Accuracy  & PGD & DF$_\infty$ &DF$_2$ & TR\\
			\midrule
			ReLU  (Base) & \textbf{94.31\%}  & 0.37\% & (1.4\%)  & (1.8\%)  &(1.3\%) \\ 
			JumpReLU (Base)        & 92.58\%           & 0.95\% & (14.3\%) & (18.5\%) &(1.9\%)  \\

			\midrule 
			\rowcolor{graytable}
			ReLU (Robust)& 93.72\%          & 60.43\%           & (6.4\%)           & (7.5\%)           & (4.8\%)\\                  
			\rowcolor{graytable}
			JumpReLU (Robust)      & 93.01\%  & \textbf{67.89\% } & (\textbf{44.4\%}) & (\textbf{43.8\%}) & (\textbf{6.1\%})\\

			\midrule 
			\rowcolor{graytable}
			Madry~\cite{madry2017towards} 				& 87.3\%           	& 50.0\%         & -    &-            	& -  \\
			\rowcolor{graytable}
			Vanilla~\cite{buckman2018thermometer}  		& 87.16\%           & 34.71\%        & -    &-           		& -  \\
			\rowcolor{graytable}
			One-hot~\cite{buckman2018thermometer}  		& 92.19\%           & 58.96\%        & -    &-           		& -  \\
			\rowcolor{graytable}			
			Thermo~\cite{buckman2018thermometer} 	& 92.32\%           & 65.67\%        & -        &-        	& -  \\
			\bottomrule 
	\end{tabular}}
	\caption{Results for WideResNet (CIFAR10); $\kappa = 0.07$.}
    \label{tab:whitebox_wide}
	\end{subtable}

	\begin{subtable}{1\textwidth}
	\centering
	\scalebox{0.99}{
		\begin{tabular}{lcccccccccc} \toprule
			Model         & Accuracy & PGD  & DF$_\infty$  &DF$_2$  &TR\\
			\midrule
			ReLU (Base) & \textbf{92.07\%} & 0.74\% & (0.7\%) & (0.9\%) &(0.7\%)\\
			JumpReLU (Base)    & 91.10\%          & 0.92\% & (5.3\%) & (6.8\%) &(1.0\%) \\

			\midrule 
			\rowcolor{graytable}
			ReLU (Robust) & 91.69\% & 53.98\%          & (4.7\%)           & (5.3\%)           &(4.1\%)\\                  
			\rowcolor{graytable}
			JumpReLU (Robust)      & 90.12\% & \textbf{59.66\%} & (\textbf{62.6\%}) & (\textbf{51.4\%}) &(\textbf{4.9\%})  \\

			\bottomrule 
	\end{tabular}}
	\caption{Results for MobileNetV2 (CIFAR10); $\kappa = 0.06$.}
    \label{tab:whitebox_mobile}
	\end{subtable}
	
\end{table}

\subsubsection{White-box attack scenario}

We next consider the more challenging white-box attack scenario. Here, the adversary has full knowledge about the model under attack, and it can access their gradients. This is the more important scenario in practice, where it is highly likely that the attacker has access to the model. 

Table~\ref{tab:whitebox} summarizes the results for the different datasets and architectures under consideration. 
Again, we see some considerable improvements for the retrofitted models---especially, the retrofitted robustly trained WideResNet (Tab.~\ref{tab:whitebox_wide}) and MobileNetV2 (Tab.~\ref{tab:whitebox_mobile}) excel. 
The performance of JumpReLU is even competitive in comparison to more sophisticated techniques such as one-hot and thermometer encoding (the authors provide only scores for the FGSM and PGD attack method)~\cite{buckman2018thermometer}.
In case of the PGD white-box attack, our retrofitted model (robust) achieves $94.36\%$ accuracy for MNIST  (Tab.~\ref{tab:whitebox_mnist}), whereas one-hot encoding achieves only $93.66\%$. 
The defense performance is also competitive for the WideResNet, where we achieve about $67.89\%$ accuracy compared to the thermometer method which achieves $65.67\%$.
Note that Buckman \etal~\cite{buckman2018thermometer} also present results for models trained with $100\%$ trained adversarial examples which outperform the results shown here. Nevertheless, these models have a lower accuracy for clean data.

Again, we want to stress the fact that JumpReLU does not requires that the model is re-trained from scratch. 
We can simply select a suitable jump value $\kappa$ during the validation stage. 
The choice of the jump value depends thereby on the desired trade-off between accuracy and robustness, \ie, large jump values improve the robustness, while decreasing the accuracy on clean examples. 
We also considered comparing with the bounded ReLU method~\cite{zantedeschi2017efficient}, but our preliminary results showed a poor performance of this defense method. A weak performance of the bounded ReLU is also reported by~\cite{carlini2017magnet}.

The adversarially trained (robust) models provide a good defense against the PGD attack (Appendix~\ref{app:pgd} contextualizes how the JumpReLU forces the PGD attack method to use more iterations in order to craft adversarial attacks).
Yet, Deep Fool is able to fool all instances in the test set using only $7$ iterations, and TR using $1000$ iterations. 
On first glance, this performance seems to be undesirable. 
We can see, however, that Deep Fool requires substantially increased average minimum perturbations in order to achieve such a high fool rate.
The numbers in parentheses in Table~\ref{tab:whitebox} indicate the average minimum perturbations which are needed to achieve a nearly $100$ percent fooling rate. 
These numbers provide a measure for the empirical robustness of the model, which we compute by using the following plug-in estimator
\begin{equation}
\tilde{\rho}_F \, :=  \, \frac{1}{n} \sum_i^n \frac{\|x_i - \tilde{x}_i\|_p}{\|x_i\|_p},
\end{equation}
with $\tilde{x}_i = x_i + \Delta x_i$. Here, we compute the relative perturbations, rather than absolute perturbations. This is because the relative measure provides a more intuitive interpretation, \ie, the numbers reflecting the average percentage of changed information in the adversarial examples.

The numbers show that the retrofitted models feature an improved robustness, while maintaining a ``good'' predictive accuracy for clean examples. For MNIST, the noise levels need to be increased by a factor of about $2$ in order to achieve a $100$ percent fooling rate. Here, we set the jump value to $\kappa=1.0$. 
For CIFAR10, we achieve a stellar performance of resilience to the Deep Fool attacks, \ie, the noise levels are required to be increased by a factor of $3$ to $7$ to achieve a successful attack.  

Clearly, we can see that the TR method is a stronger attack than Deep Fool. However, we are still able to achieve an improved resilience to this strong attack. For instance, the TR attack requires average minimum perturbations of about $6.1\%$ to attack the retrofitted robust WideResNet (Tab.~\ref{tab:whitebox_wide}) and about $4.9\%$ to attack the retrofitted robust MobileNetV2 (Tab.~\ref{tab:whitebox_mobile}). These high levels of perturbations are critical in a sense that they are not any longer imperceptible for humans, i.e., they render the attack less useful in practice.

%. However, no statistical test can tell you whether the effect is large enough to be important in your field of study.

\subsubsection{Performance trade-offs}

\begin{figure}[!b]
	\centering
	\begin{subfigure}[t]{0.49\textwidth}
		\includegraphics[width=1\textwidth]{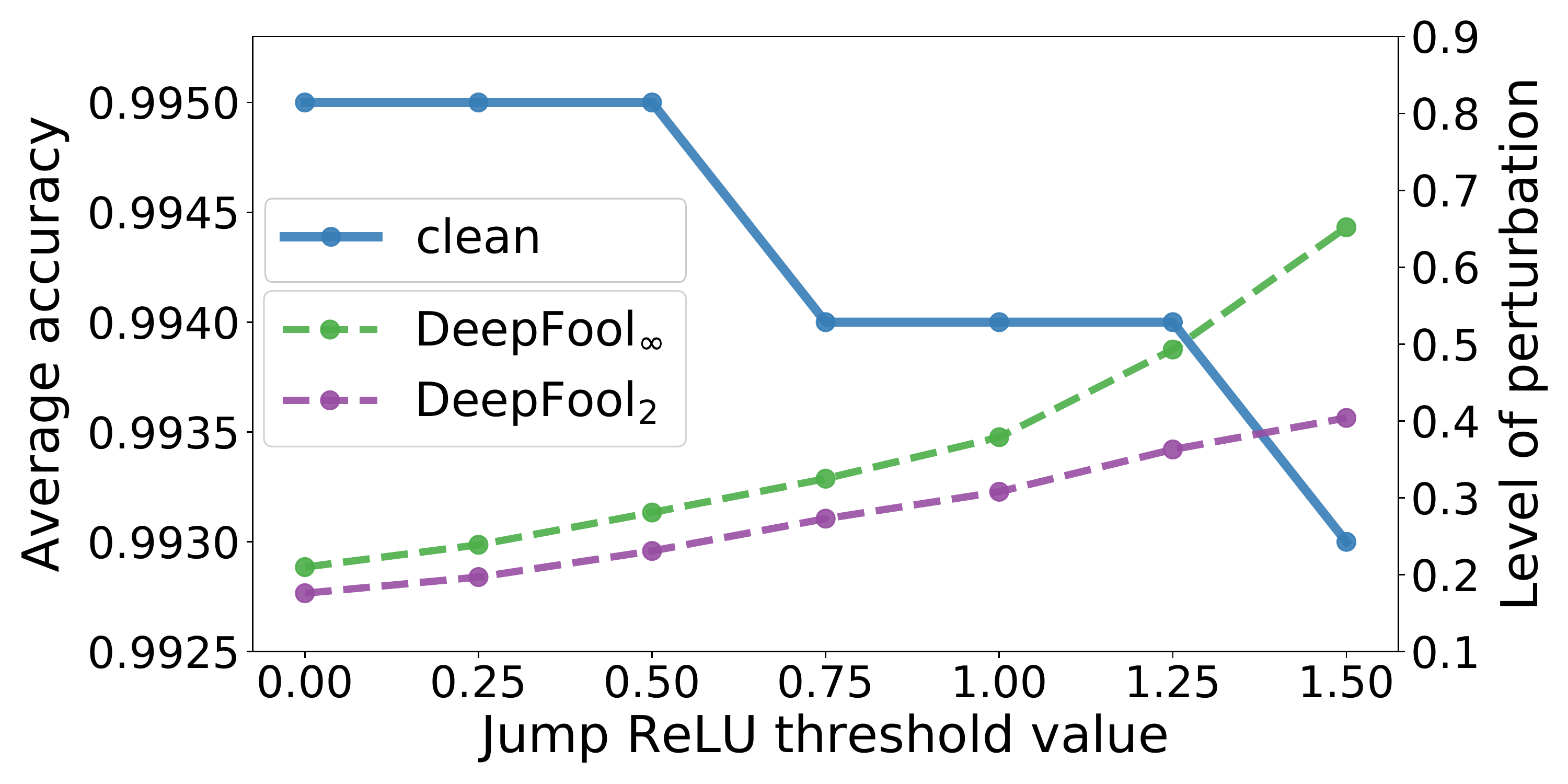}
		\caption{LeNetLike network (MNIST)}
	\end{subfigure}
	~
	\begin{subfigure}[t]{0.49\textwidth}
		\includegraphics[width=1\textwidth]{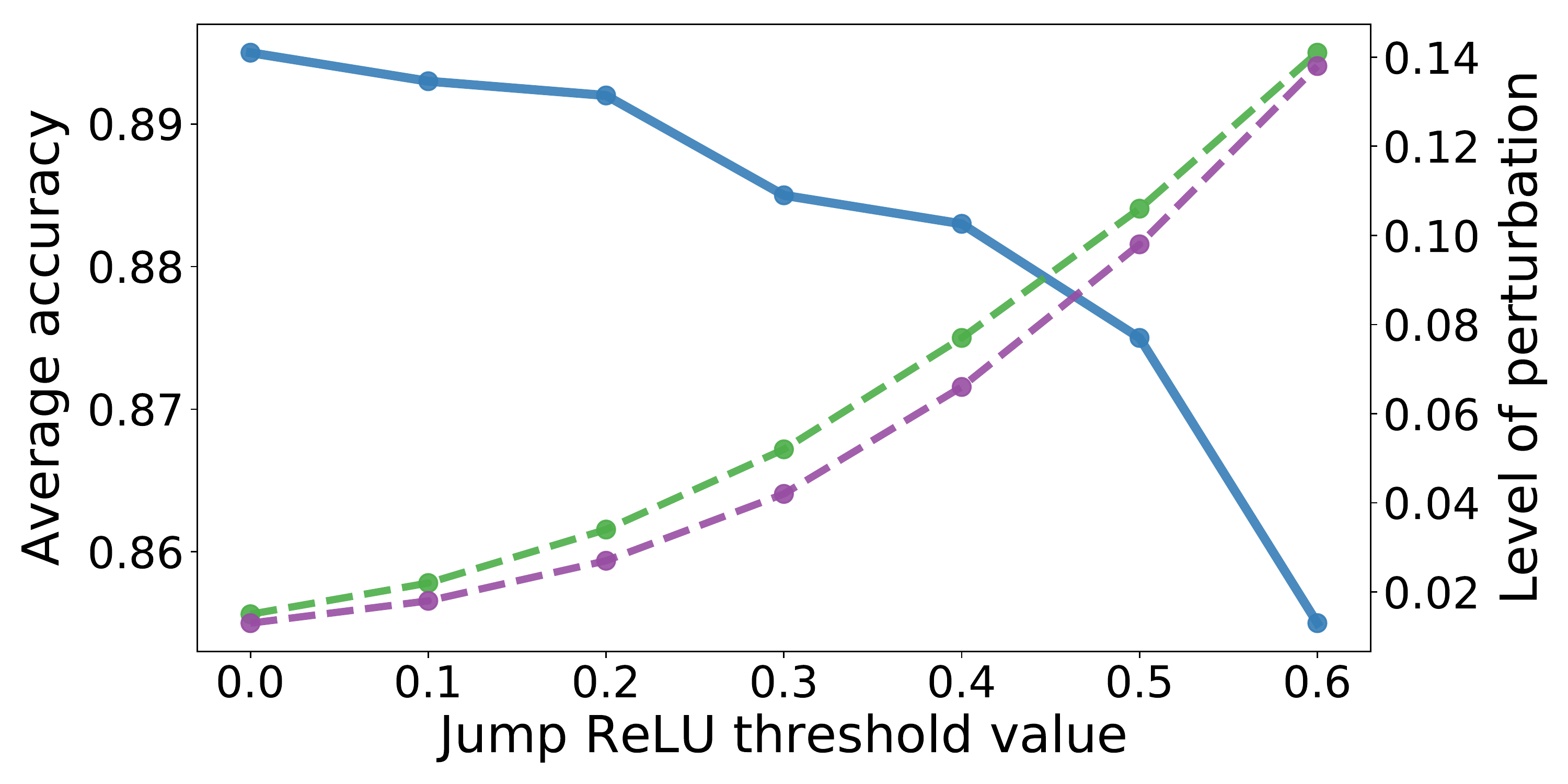}
		\caption{AlexLike network (CIFAR10)}
	\end{subfigure}
\vspace{+0.4cm}

	\begin{subfigure}[t]{0.49\textwidth}
	\includegraphics[width=1\textwidth]{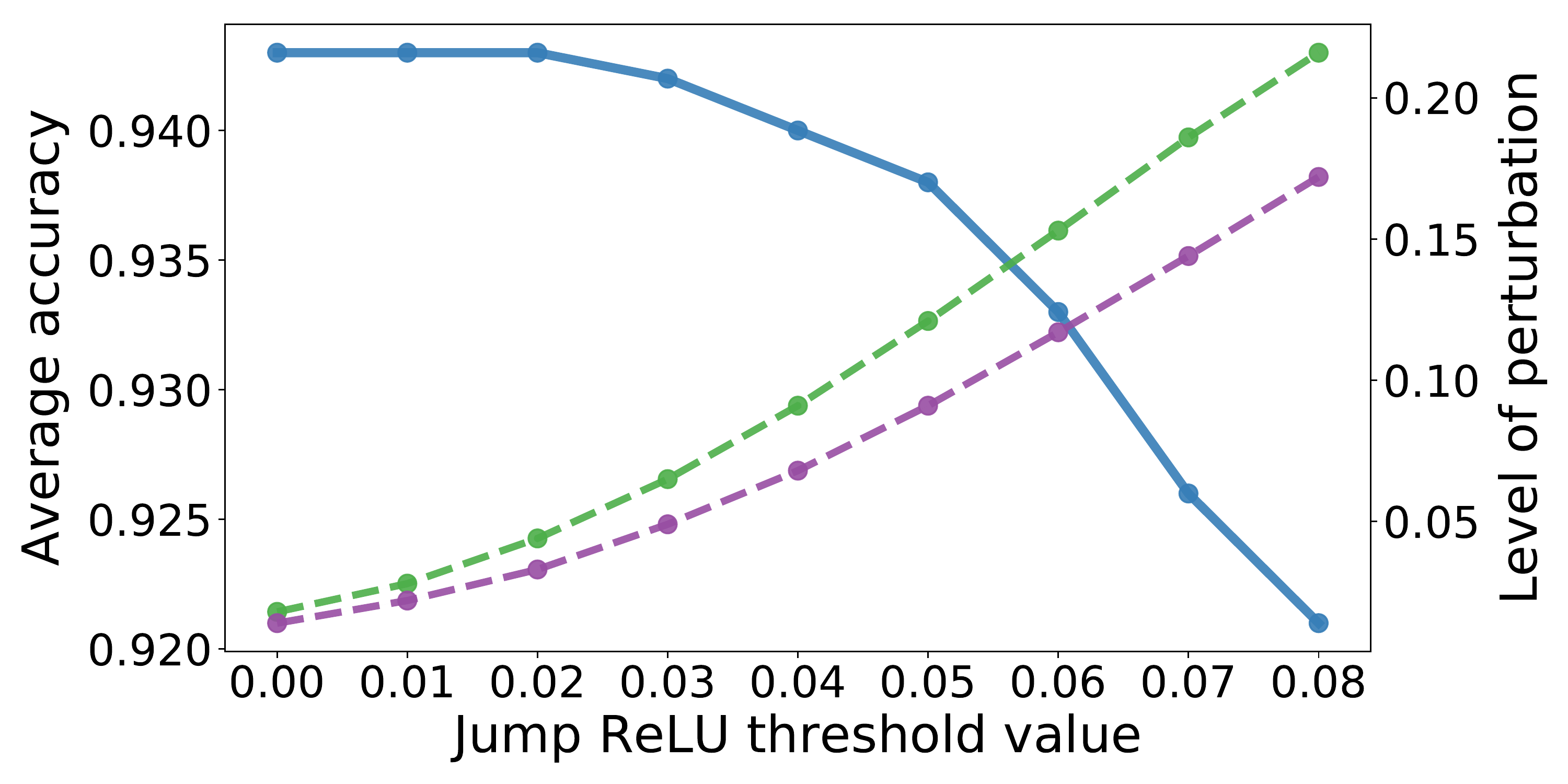}
	\caption{WideResNet (CIFAR10)}
	\end{subfigure}
	~
	\begin{subfigure}[t]{0.49\textwidth}
	\includegraphics[width=1\textwidth]{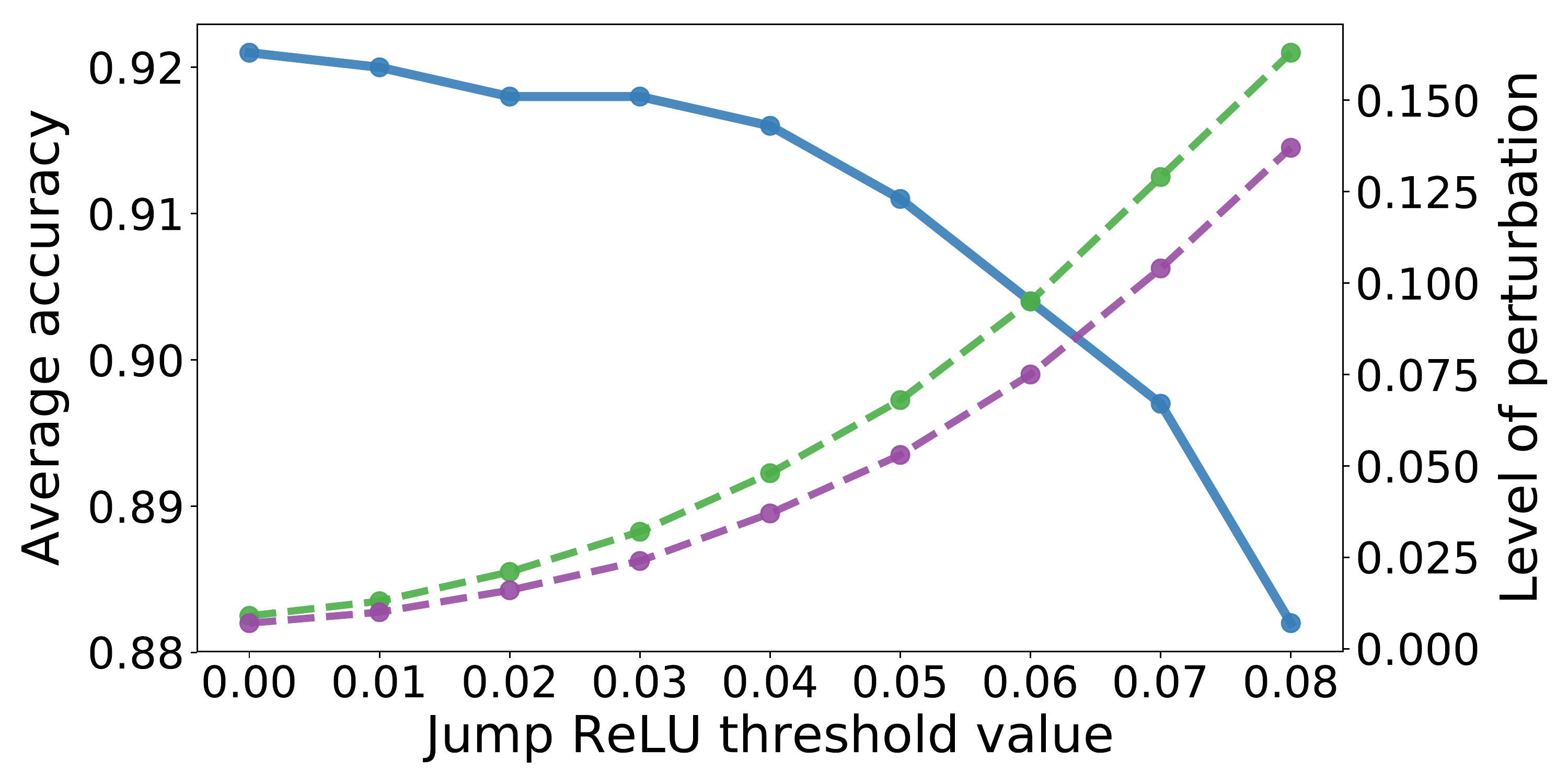}
	\caption{MobileNetV2 (CIFAR10)}
	\end{subfigure}	
	
	\caption{JumpReLU performance trade-offs for MNIST and CIFAR10. The left axis shows the average predictive accuracy for clean examples for varying values of the jump value. The right axis shows the average minimum perturbations required to construct adversarial examples which achieve a nearly $100\%$ fooling rate.}
	\label{fig:performance_tradeoff}
\end{figure}

As mentioned, the JumpReLU activation function provides a trade-off between robustness and classification accuracy. 
The user can control this trade-off in a post-training stage by tuning the jump value $\kappa$, where $\kappa=0$ resembles the ReLU activation function. 
Of course, the user needs to decide how much accuracy on clean data he is willing to sacrifice in order to buy more robustness. However, this sacrifice is standard to most robustification strategies. For instance, for adversarial training one must choose the ratio between clean and adversarial examples used for training, where a higher ratio of adversarial to clean examples improves the robustness while decreasing the predictive accuracy. 
Thus, the decision of a ``good'' jump value is application dependent. 

Figure~\ref{fig:performance_tradeoff} shows this trade-off for different network architectures. 
We see that the jump value $\kappa$ is positively correlated to the level of perturbation which is required in order to achieve a $100$ percent fooling rate. 
%
%We can see that the model's average predictive accuracy is insignificantly affected by small jump values.
% 
Choosing larger jump values increase the robustness of the model, while sacrificing only a slight amount of predictive accuracy. 
It can be seen, that larger jump values only marginally effect the accuracy of the LeNetLike network on clean examples, while the other networks are more sensitive.

\subsubsection{Visual results}

The interested reader may ask whether the increased adversarial perturbations are of any practical significance. To address this question, we show some visual results which illustrate the magnitude of the effect. 
Recall the aim of the adversary is to construct unobtrusive adversarial examples. 

\begin{figure}[!t]
	\centering
\begin{minipage}{.5\textwidth}
	\centering
	\begin{subfigure}[t]{1\textwidth}
		\includegraphics[width=1\textwidth]{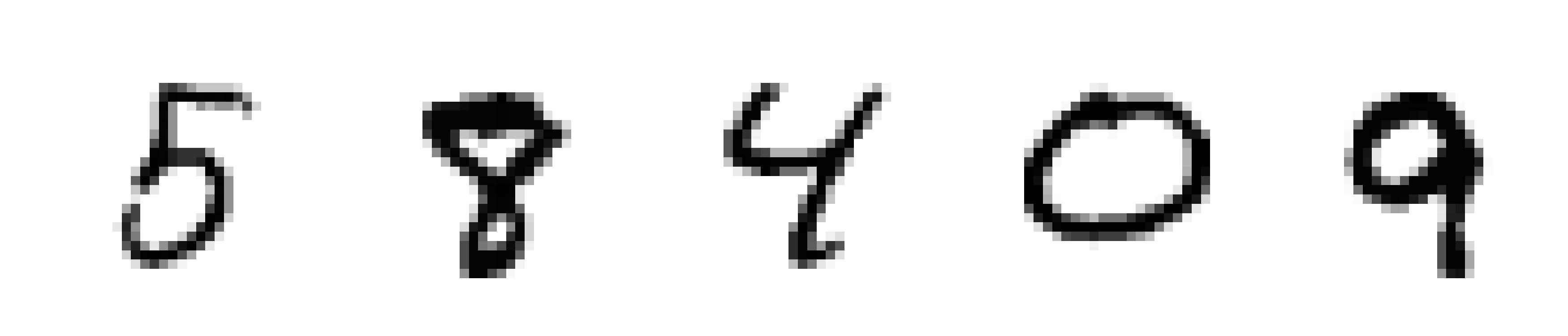}
		\vspace{-0.7cm}
		\caption{\footnotesize Clean examples which are used for training.}
	\end{subfigure}\vspace{+0.2cm}  
	
	\begin{subfigure}[t]{1\textwidth}
		\includegraphics[width=1\textwidth]{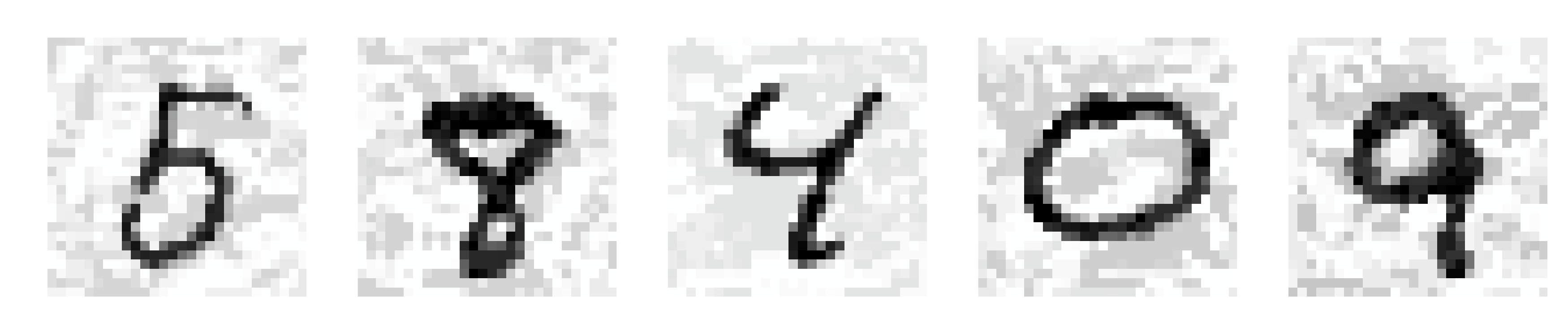}
		\vspace{-0.7cm}
		\caption{\footnotesize Adversarial examples to fool the model without defense.}
	\end{subfigure}\vspace{+0.2cm}  
	
	\begin{subfigure}[t]{1\textwidth}
		\includegraphics[width=1\textwidth]{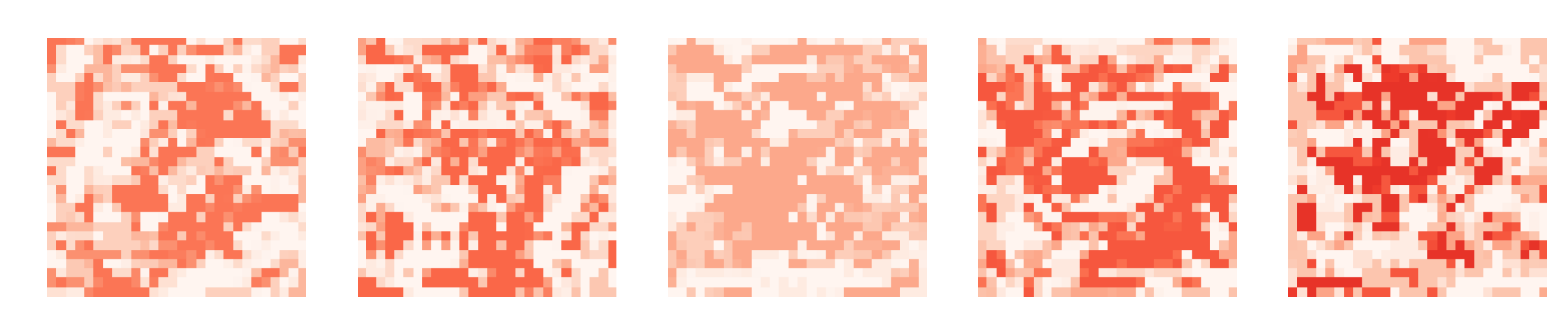}
		\vspace{-0.7cm}
		\caption{\footnotesize Perturbation patterns to fool the model without defense.}
	\end{subfigure}\vspace{+0.2cm}  
	
	\begin{subfigure}[t]{1\textwidth}
		\includegraphics[width=1\textwidth]{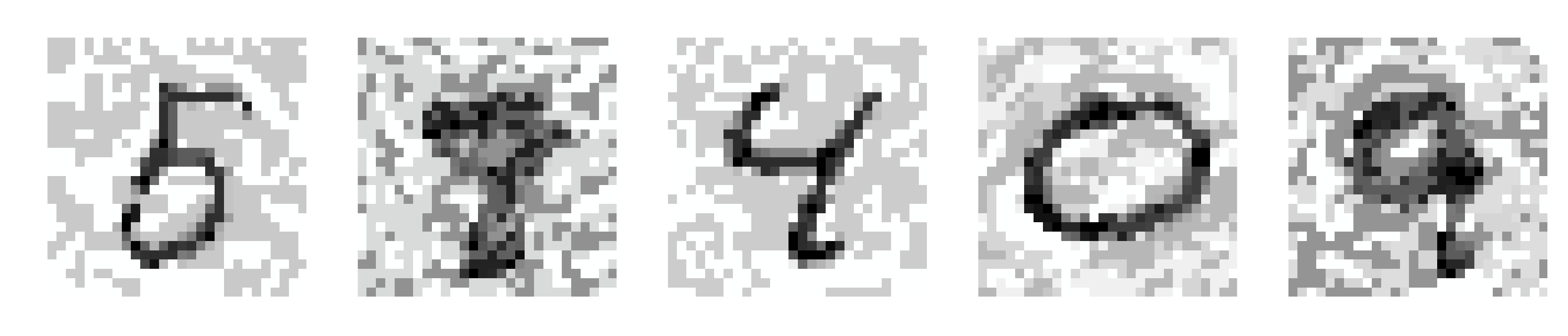}
		\vspace{-0.7cm}		
		\caption{\footnotesize Adversarial examples to fool the retrofitted model.}
	\end{subfigure}    
	
	\begin{subfigure}[t]{1\textwidth}
		\includegraphics[width=1\textwidth]{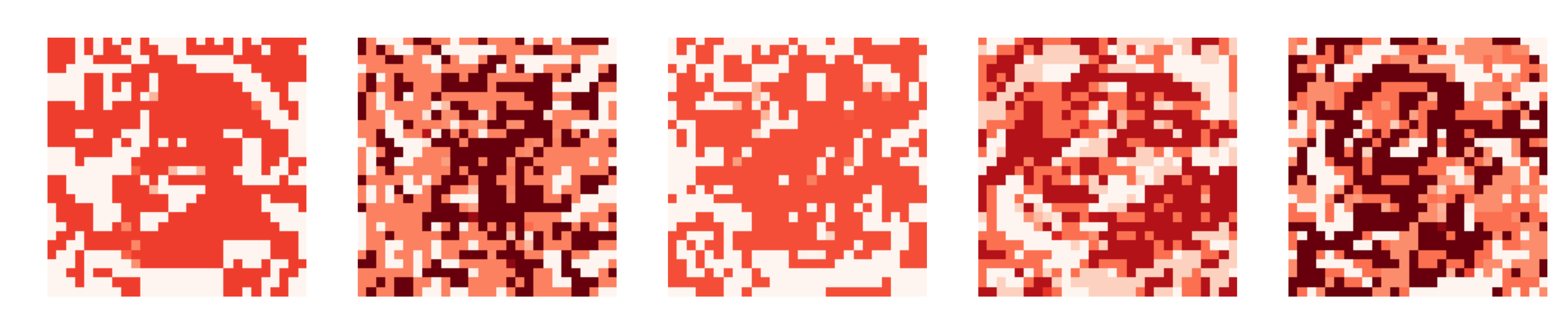}
		\vspace{-0.7cm}		
		\caption{\footnotesize Perturbation patterns to fool the retrofitted model.}
	\end{subfigure}     
	
	\caption{Visual results for MNIST to verify the effect of the JumpReLU defense strategy against the DF$_\infty$ attack. Noticeable higher levels of perturbations are required in order to successfully attack the retrofitted network. Subfigures (c) and (e) show the corresponding perturbation patterns. %Here, we use a fixed color map between $0.0$ (light red) and $1.0$ (dark red).
	}
	\label{fig:mnist_vis}
\end{minipage}%	
~
\begin{minipage}{.5\textwidth}
    %\vspace{-0.0cm}
	\begin{subfigure}[t]{1\textwidth}
		\includegraphics[width=1\textwidth]{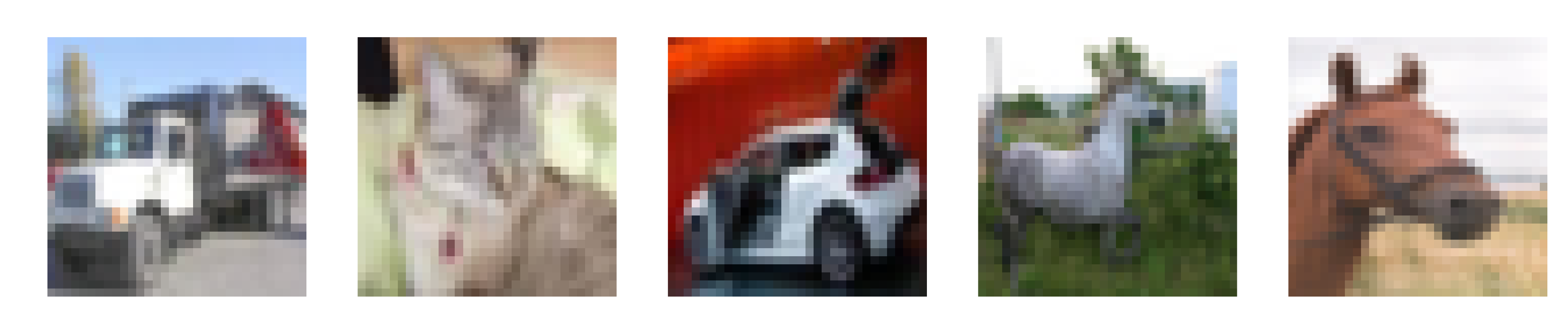}
		\vspace{-0.7cm}
		\caption{\footnotesize Clean examples which are used for training.}
	\end{subfigure}\vspace{+0.2cm} 
	
	\begin{subfigure}[t]{1\textwidth}
		\includegraphics[width=1\textwidth]{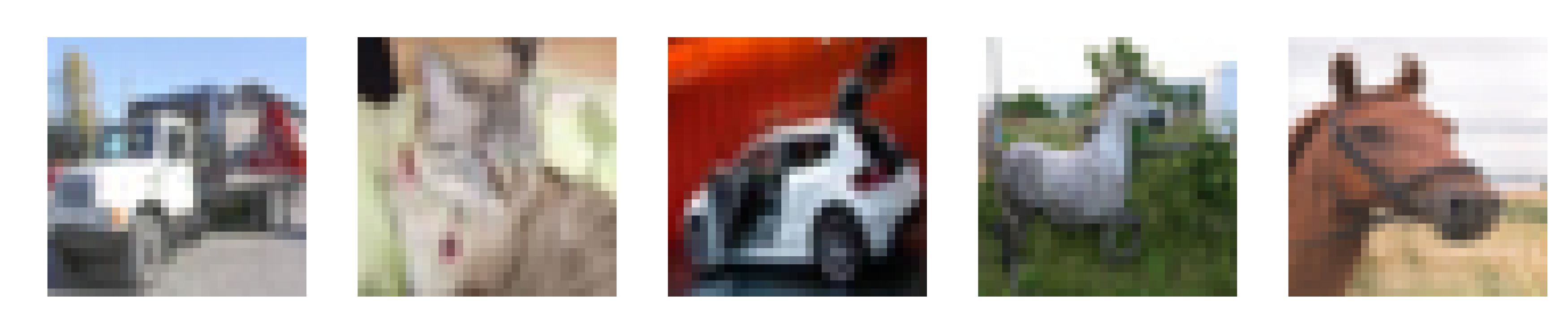}
		\vspace{-0.7cm}		
		\caption{\footnotesize Adversarial examples to fool the model without defense.}
		\label{fig:cifar_vis_relu}
	\end{subfigure}\vspace{+0.2cm}  
	
	\begin{subfigure}[t]{1\textwidth}
		\includegraphics[width=1\textwidth]{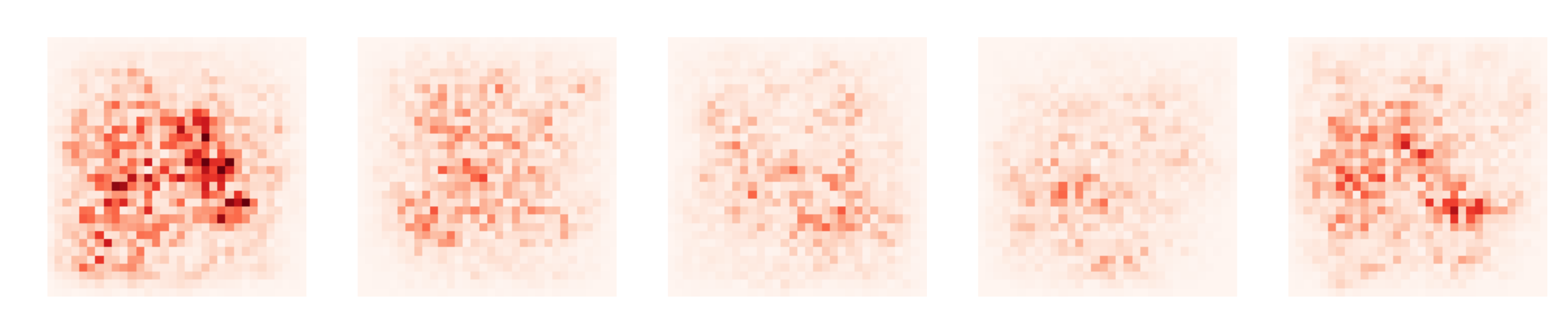}
		\vspace{-0.7cm}		
		\caption{\footnotesize Perturbation patterns to fool the model without defense.}
	\end{subfigure}\vspace{+0.2cm} 
	
	\begin{subfigure}[t]{1\textwidth}
		\includegraphics[width=1\textwidth]{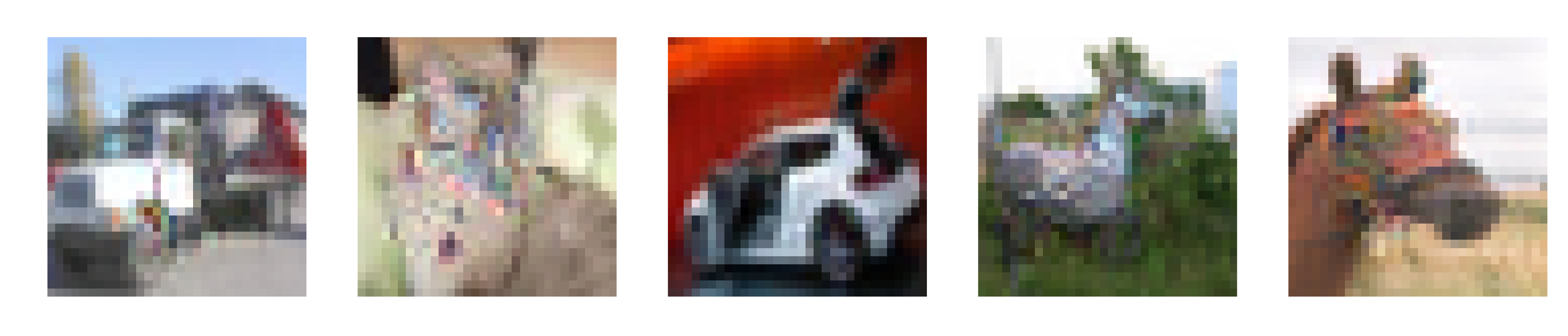}
		\vspace{-0.7cm}		
		\caption{\footnotesize Adversarial examples to fool the retrofitted model.}
		\label{fig:cifar_vis_jump}
	\end{subfigure}        
	
	\begin{subfigure}[t]{1\textwidth}
		\includegraphics[width=1\textwidth]{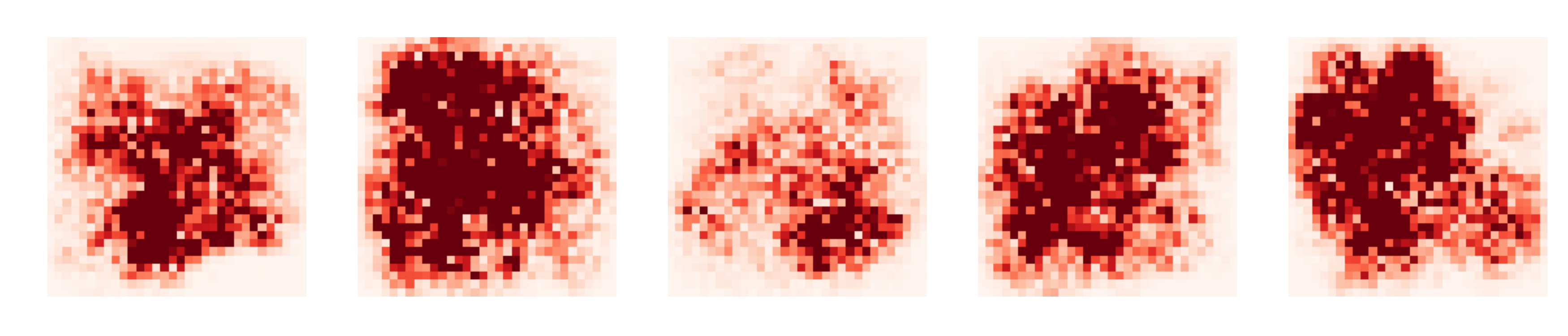}
		\vspace{-0.7cm}		
		\caption{\footnotesize Perturbation patterns to fool the retrofitted model.}
	\end{subfigure} 
	
	\caption{Visual results for CIFAR10 to verify the effect of the JumpReLU defense strategy against the DF$_\infty$ attack. By visual inspection, it can b seen that the DF$_2$ attack requires noticeable higher levels of perturbations in order to successfully attack the retrofitted network. %Here we use a fixed color map between $0.0$ (light red) and $0.2$ (dark red).
	}
	\label{fig:cifar_vis}

\end{minipage}%
\end{figure}

Figure~\ref{fig:mnist_vis} shows both clean and adversarial examples for the MNIST dataset, which are crafted by the Deep Fool algorithm. 
Clearly, the adversarial examples which are needed to fool the retrofitted LeNetLike network are visually distinct from those examples which are sufficient to fool the unprotected model.
We also show the corresponding perturbation patterns, \ie, the absolute pixel-wise difference between the clean and adversarial examples, to better illustrate the difference. 
Note that we use a ``reds'' color scheme here: white indicates no perturbations, light red indicates very small perturbations, dark red indicates large perturbations.

Next, Figure~\ref{fig:cifar_vis} shows visual results for the CIFAR10 dataset. It is well known that models for this dataset are highly vulnerable, \ie, very small perturbations $\Delta x$ are already sufficient for a successful attack. 
Indeed, the minimal perturbations which are needed to fool the unprotected network (here we show results for the AlexLike network) are nearly imperceptible by visual inspection. 
In contrast, the crafted adversarial examples to attack the retrofitted model show distinct perturbation patterns, and one can recognize that the examples were altered. 
Note the example we show here correspond to the baseline AlexLike network. 

In summary, the visual results put the previously presented relative noise levels into perspective, and they show that average minimum perturbations of about $5\%$ to $10\%$ are lucid. Thus, it can be concluded that the JumpReLU is an effective strategy for improving the model robustness.

\subsubsection{Adversarial detection}

As a proof-of-concept, we demonstrate that the increased minimum perturbations, which are required to attack the retrofitted model can help to improve the discrimination power of add-on adversarial detectors.
While for humans adversarial perturbations are often visually imperceptible, add-on detectors aim to discriminate between clean and adversarial examples using inputs from intermediate feature representations of a model. 
Indeed, these specifically trained detectors have been shown to be highly effective for detecting adversarial examples~\cite{metzen2017detecting,grosse2017statistical,lu2017safetynet}. Yet, there is also work which shows that adversarial detectors can be fooled (bypassed) if the attacker is aware of their presence~\cite{carlini2017adversarial}. However, such specific attacks require to be more sophisticated than the commonly used attack methods.

We follow the work by Ma \etal~\cite{ma2018characterizing}, who use the idea of Local Intrinsic Dimensionality (LID) to characterize adversarial subspaces. The idea is that clean and adversarial examples show distinct patterns so that the LID characteristics allow to discriminate between such examples. 

Intuitively, adversarial examples which show increased perturbation patterns should feature more extreme LID characteristics. Hence, a potential application of the JumpReLU is to combine it with an LID based detector. Table~\ref{tab:AUC_scores} shows the area under a receiver operating characteristic curve (AUC) as a measure for the discriminate power between clean and adversarial examples. Indeed, the results show that the combination with JumpReLU improves the detection performance for CIFAR10.

\begin{table}[!b]
\caption{AUC scores as measure of the discrimination power between clean and adversarial examples using LID characteristics. Here we compare ReLU and JumpReLU.}\label{tab:AUC_scores}
\centering
\scalebox{0.99}{
\begin{tabular}{lcccccccccc} \toprule
    Model        & PGD     & DF$_\infty$        &DF$_2$      &TR\\
    \midrule
LID + ReLU      &72.54  &73.41  &72.93 & 72.47 \\
LID + JumpReLU  & \textbf{74.25}  & \textbf{78.24}  & \textbf{75.84} & \textbf{74.71} \\
    \bottomrule 
\end{tabular}}
\end{table}

\section{Conclusion}

We have proposed a new activation function---the \textit{JumpReLU} function---which, when used in place of a ReLU in an already pre-trained model, leads to a trade-off between  predictive  accuracy  and  robustness.
This  trade-off is controlled by a parameter, the jump size, which can be tuned during the validation stage.
That is, no additional training of the pre-trained model is needed when the JumpReLU function is used.
(Of course, if one wanted to perform additional expensive training, then one could do so.)
Our experimental results show that this simple and inexpensive strategy improves the resilience to adversarial attacks of previously-trained networks.
Appendix~\ref{app:rand} explores extension of the JumpReLU. Randomness as a resource to improve model robustness has been demonstrated before within the defense literature. Motivated by this observation, we introduce the randomized JumpReLU and show that a small amount of randomness can help to improve the model robustness even further.

Limitations of our approach are standard for current adversarial defense methods, in that stand-alone methods do not guarantee a holistic protection and that sufficiently high levels of perturbation will be able to break the defense. 
That being said, JumpReLU can easily be used as a stand-alone approach to ``retrofit'' previously-trained networks, improving their robustness, and it can also be used to support other more complex defense strategies.
%
%Indeed, we showed that increased perturbations actually affects the characteristics of the inner filter output, which leads to better detection accuracy of adversarial examples. 

% Although the adversarial perturbations of images are invisible to humans, we showed that the effect of the perturbations actually significantly affects the characteristics of the inner filter output. 
% %Indeed, every filter has a characteristic kernel density shape. 
% %
% A targeted attack can be performed by crafting an adversarial example which leads to filter activation so that the filter outputs resemble the shape of the kernel density of the target class.
% %
% We deomonstrated that JumpReLU is able to reduce this effect and can improve the resilience of already deployed networks. 
% %
% This is simply achieved by replacing ReLU by the JumpReLU activation function with a threshold value $\kappa > 0$. 
% %

% Future research directions are to investigate the the effect of JumpReLU during the training stage as well as to transfer the idea to other activation functions. 

\clearpage

\appendix

\section{A second look to the gray-box attack properties of the JumpReLU}\label{app:second_graybox}

\begin{figure}[!b]
	\centering
	%\vspace{-.20in}
	\begin{subfigure}[t]{0.45\textwidth}
		\includegraphics[width=1\textwidth]{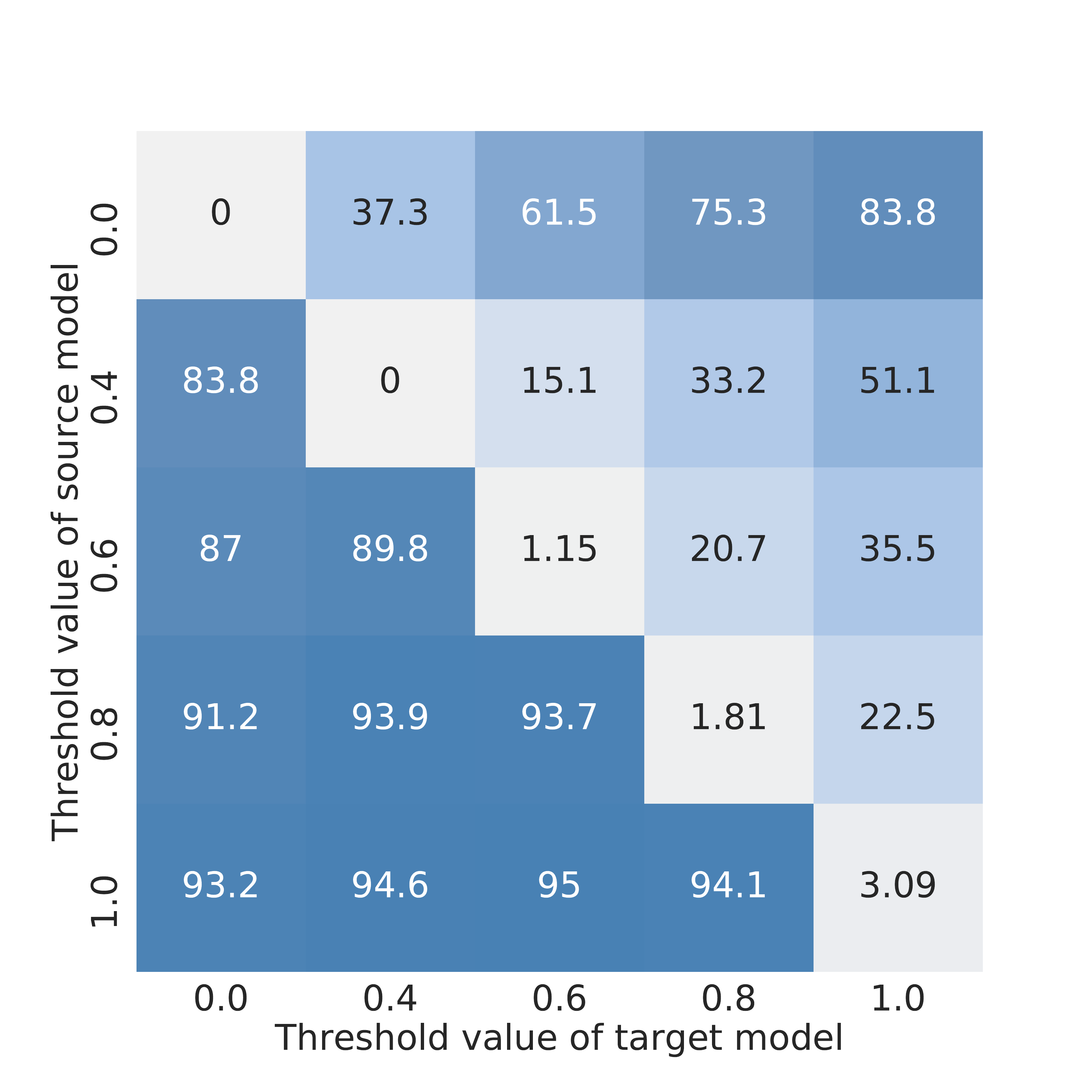}
		\caption{Transferability for MNIST.}
	\end{subfigure}
	~
	%\vspace{+.10in}
	\begin{subfigure}[t]{0.45\textwidth}
		\includegraphics[width=1\textwidth]{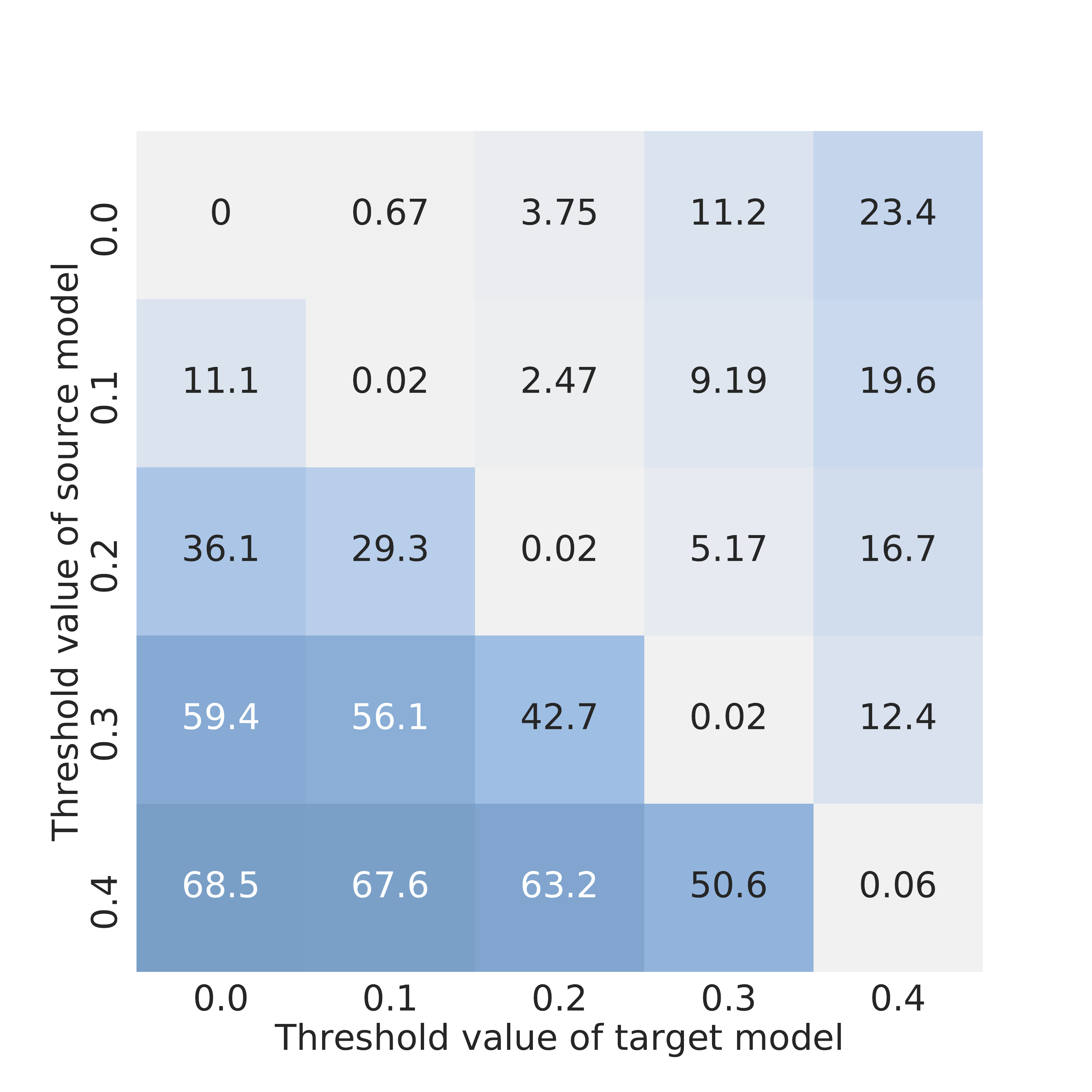}
		\caption{Transferability for CIFAR10.}
	\end{subfigure}
	
	\caption{Gray-box attack matrix for different jump values. Each cell $(i,j)$ indicates the predictive accuracy of a model retrofitted with the jump value $j$ (target), which is being attack by using adversarial examples generated by a model with jump value $i$ (source). Higher cell values indicate better robustness. }
	\label{fig:transferability}
\end{figure}
%

%The phenomena of transferability of adversarial examples has recently attracted some attention~\cite{papernot2016transferability,carlini2017towards}. It has been demonstrated that adversarial examples generated by a specific model can also be used to fool other  machine learning techniques~\cite{papernot2016transferability}. For instance, adversarial  examples crafted to fool a neural network can have strong cross-technique transferability properties and can be used to attack support vector machines or decision trees.
%
%Thus, this intriguing property lead to the possibility of performing efficient black-box attacks. 

We present an extended set of results for the gray-box attack
scenario. Specifically, we study the situation where the adversary has  full  access  to a retrofitted model (which has a fixed jump value) in order to construct adversarial examples, but the adversary has no information about the jump value of the target network during inference time.

Here, the adversarial examples are crafted by using the projected gradient decent (PGD) attack method. Figure~\ref{fig:transferability} shows the efficiency of a non-targeted attack on networks using different jump values. Note, we run the attack with a large number of iterations, enough so that the crafted adversarial examples achieve a nearly 100 percent fool rate for the source model. 

We see that the attack is unidirectional, \ie, adversarial examples crafted by using source models which have a low jump value can be used to attack  models which have a higher jump value. However, retrofitted models which have a low jump value are resilient to adversarial examples generated by source models which have a large jump value.
Thus, one could robustify the network by using a large jump size $\kappa$ for evaluating the gradient, while using a smaller jump size for inference. Of course, this is a somewhat pathological setup, designed to illustrate and validate properties of the method, yet these results reveal some interesting behavioral properties of the JumpReLU.

\section{Accuracy vs number of iterations}\label{app:pgd}

Iterative attack methods can be computational demanding if a large number of iterations is required to craft strong adversarial examples. Of course, it is an easy task to break any defense with unlimited time and computational resources. However, it is the aim of an attacker to design efficient attack strategies (\ie, fast generation of examples which have minimal perturbations), while the defender aims to make models more robust to these attacks (\ie, force the attacker to increase the average minimal perturbations which are needed to fool the model).

\begin{figure}[!b]
	\centering
	%\vspace{-.20in}
	\begin{subfigure}[t]{0.43\textwidth}
		\includegraphics[width=1\textwidth]{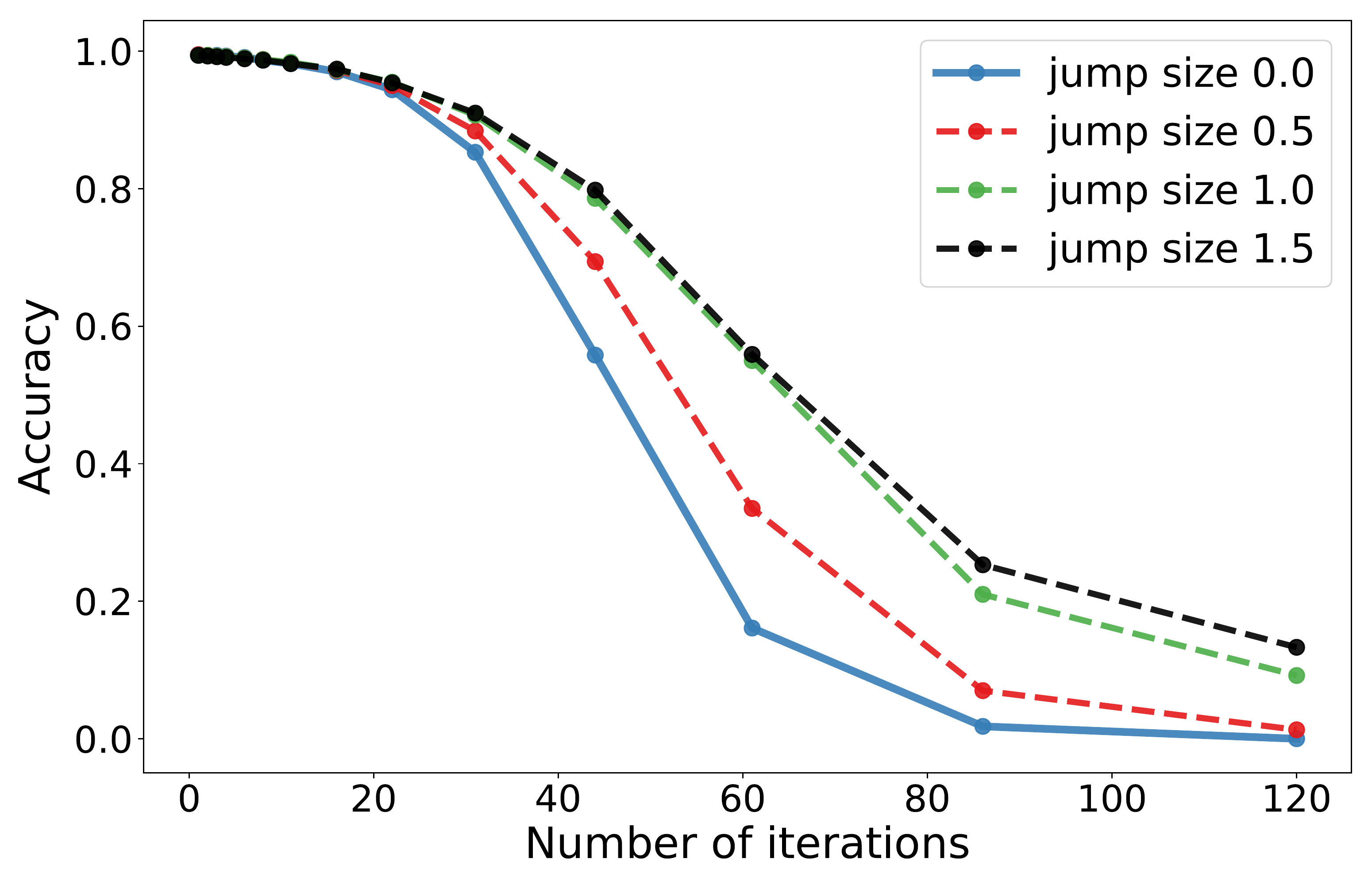}
		\caption{LeNetLike network (MNIST).}
	\end{subfigure}
	~
	%\vspace{+.10in}
	\begin{subfigure}[t]{0.43\textwidth}
		\includegraphics[width=1\textwidth]{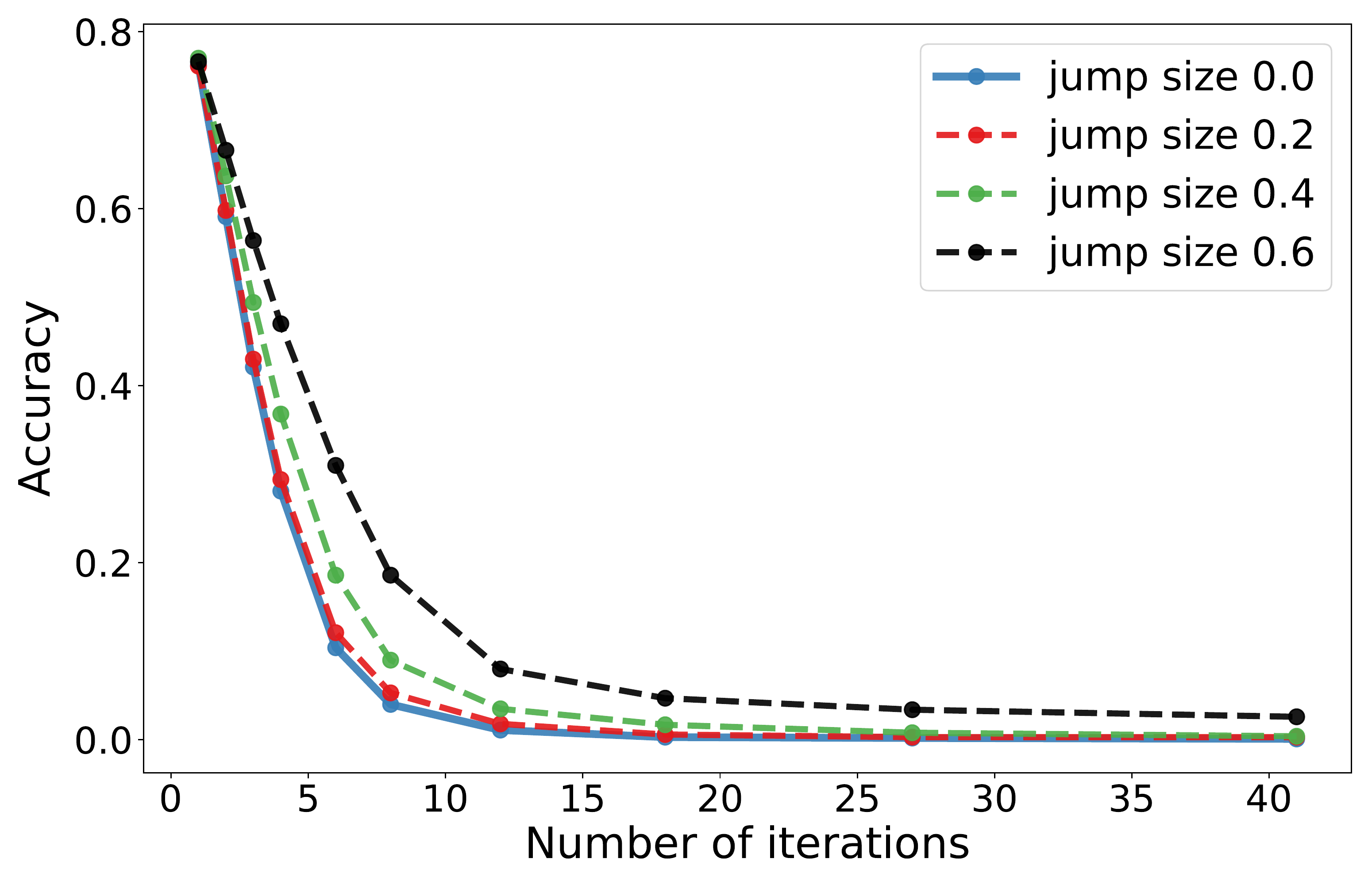}
		\caption{AlexLike (CIFAR10).}
	\end{subfigure}
	
	\begin{subfigure}[t]{0.43\textwidth}
		\includegraphics[width=1\textwidth]{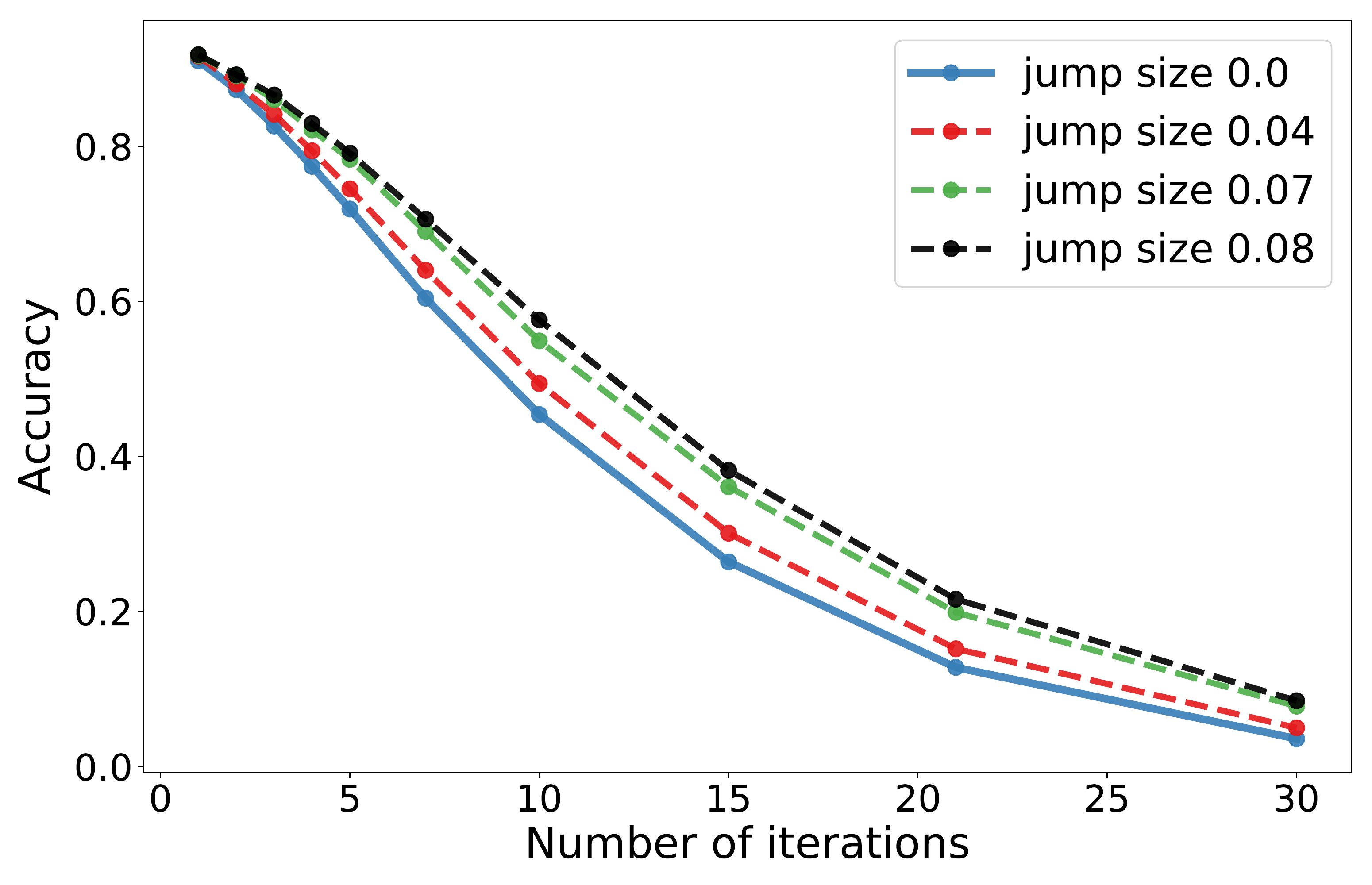}
		\caption{Robust WideResNet (CIFAR10).}
	\end{subfigure}
	~
	%\vspace{+.10in}
	\begin{subfigure}[t]{0.43\textwidth}
		\includegraphics[width=1\textwidth]{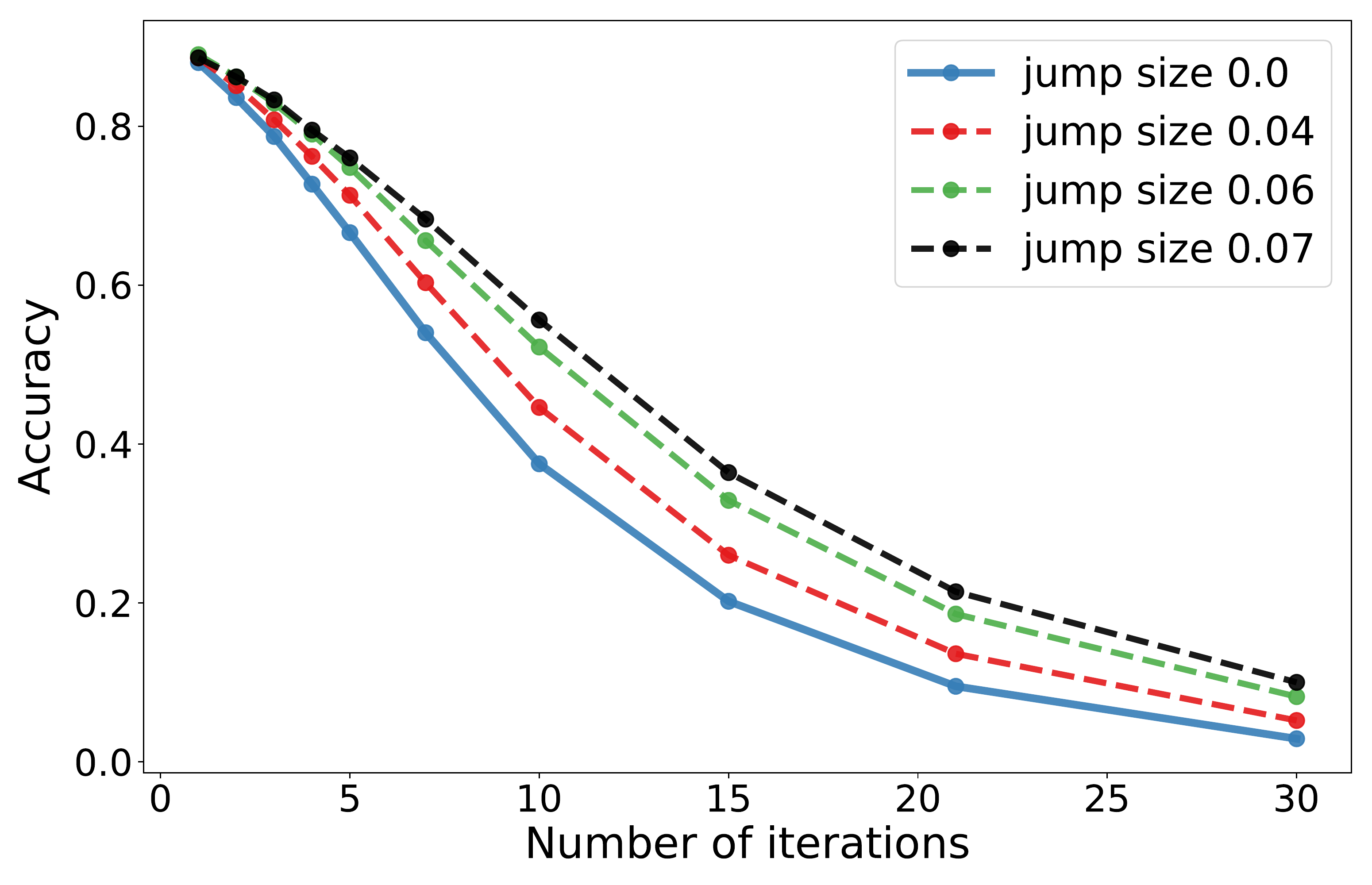}
		\caption{Robust MobileNetV2 (CIFAR10).}
	\end{subfigure}    
	
	\caption{Strength of the PGD attack for increasing numbers of iterations. It can be seen, that the PGD method requires a large number of iterations to craft strong adversarial examples. The JumpReLU increases the model robustness, i.e., the fooling rate is reduced for a fixed number of iterations.}
	\label{fig:pgd_iter}
\end{figure}

Figure~\ref{fig:pgd_iter} contextualizes the accuracy vs the number of iterations for the PGD attack. Attacking the retrofitted model requires a larger number of iterations in order to achieve the same fool rate as for the unprotected network. 
This is important, because a large number of iterations requires more computational resources as well as it increases the computational time.
To put the numbers into perspective, it takes about $4$ minutes to run $7$ iterations to attack the unprotected WideResNet. In contrast, it takes about $5$ minutes to run $7$ iterations to attack the retrofitted model. 

\section{Randomized JumpReLU}\label{app:rand}

Several state-of-the-art defense strategies rely on randomness as a resource for improving model robustness. Here, we explore whether a randomized version of the JumpReLU can help to further improve the model robustness. 

More concretely, the randomized JumpReLU selects a random $\kappa$ in a specified range for every forward pass. The underlying idea is that this approach leads to obfuscated gradients. It has impressively demonstrated that obfuscated gradients do not guarantee safety~\cite{athalye2018obfuscated}. 
Nevertheless, our aim is to evaluate whether the randomized JumpReLU leads to increased average minimal perturbations.
Table~\ref{tab:whitebox_random_jumprelu} shows the results for the white-box attack scenario. For comparison we show here also the results for the deterministic JumpReLU. Here, we chose the jump value so that the retrofitted models, using the deterministic and randomized JumpReLU, have roughly the similar accuracy for clean data. 

First, we note that the average minimal perturbations are increased in all situations, especially those for the TR attack method. 
For instance, the TR attack needs to increase the average minimal perturbations from $6.1\%$ to $18.1\%$ to attack the robust WideResNet, yet it achieves only a fool rate of $70.73\%$. We see a similar behavior for the MobileNetV2 architecture. There are substantial gains in terms of the model robustness, and this renders the TR attack useless, since adversarial examples featuring such large perturbation patterns are easy to detect. 
Second, we see that in many instances the different attack methods fail to achieve a nearly 100 percent fool rate despite the increased perturbations.

This leads to the conclusion that randomness can indeed help to improve the robustness. However, the drawback is that this approach requires a second tuning parameter. That is, because we sample $\kappa$ from a uniform distribution with support $\kappa \in [a,b]$. For our experiments, we simply set $a := 0.6\cdot b$. We have not explored different settings and leave this open as a future research direction.

%Further, we show the performance of the randomized JumpReLU for two different sets of jump values $\kappa$. The results show the performance trade-offs between accuracy and robustness. Increasing the jump value improves the model robustness, while sacrificing slightly more accuracy on clean data. 

\begin{table}[!t]
	%\vspace{-0.4cm}	
	\caption{Summary of results for white-box attacks with \textbf{randomized JumpReLU}. Here (D) denotes the deterministic and (R) denotes the randomized JumpReLU.
		The numbers indicate the accuracy, \ie, the percentage of correctly classified instances (higher numbers indicate better robustness). In addition, we show the average minimum perturbations in parentheses. The best performance is highlighted in bold letters.}
	\label{tab:whitebox_random_jumprelu}
	\setlength\tabcolsep{3.2pt}

	\begin{subtable}{1\textwidth}
	\centering
	\scalebox{0.99}{
		\begin{tabular}{lcccccccccc} \toprule
			Model         & $\kappa$     & Accuracy    & PGD  & DF$_\infty$  &DF$_2$  &TR\\
			\midrule
			ReLU  (Base) & 0.00 & 99.55\%          & 66.69\% & 0.00\% (17.9\%) & 0.00\%  (21.8\%) & 0.00\%  (18.9\%) \\            
			JumpReLU (D)  & 1.00 & 99.53\%          & 83.21\% & 0.00\%  (34.1\%) & 0.00\%  (44.9\%) & 0.00\%  (25.0\%) \\
			
			JumpReLU (R) & 1.00 & \textbf{99.57\%} & 83.49\% & 5.37\%  (36.2\%) & 2.64\%  (47.6\%) & 9.61\%   (37.0\%) \\
			
			%JumpReLU (R) & 1.25 & 99.48\%          & {86.62\%} & \textbf8.70\%} ({47.1\%}) & \textbf{5.40\%} ({70.0\%}) & \textbf{11.56\%} ({45.1\%}) \\

			\midrule 
			\rowcolor{graytable}
			ReLU  (Robust)  & 0.00 & 99.50\%  & 91.39\%          & 0.00\% (28.4\% )          & 0.0\% (31.4\% )          & 0.00\% (24.7\%)\\ 
			\rowcolor{graytable}
			JumpReLU (D) &	1.00  & 99.47\%  & 94.36\%          & 0.00\% (46.6\%)           & 0.00\% ({53.3\%})  & 0.00\% (32.8\%) \\
			\rowcolor{graytable}
			JumpReLU (R)  &  1.00  & 99.47\%  & \textbf{95.17\%} & \textbf{5.89\%} (\textbf{51.0\%})  &  \textbf{1.39\%}  (\textbf{52.8\%})           & \textbf{8.08\%} (\textbf{44.8\%})\\
			
			%\rowcolor{graytable}
			%JumpReLU (R)  &  1.25  & 99.44\%  & \textbf{95.53\%} & {8.11\%} (\textbf{58.8\%})  & {3.44\%} (\textbf{70.4\%})           & {10.08\%} (\textbf{54.8\%})\\

			\bottomrule 
	\end{tabular}}
	\caption{Results for LeNetLike network (MNIST).}
    \label{tab:randrelu_mnist}
	\end{subtable}
	
    \begin{subtable}{1\textwidth}	
	\centering
	\scalebox{0.99}{
		\begin{tabular}{lcccccccccc} \toprule
			Model       &  $\kappa$   	& Accuracy     & PGD   & DF$_\infty$ &DF$_2$ &TR\\
			\midrule
			ReLU  (Base)  & 0.00 & \textbf{89.46\%}	& 6.38\%  & 0.00\%  (1.2\%)	 & 0.00\%  (1.5\%)  & 0.00\%  (1.3\%) \\                
			JumpReLU (D)  & 0.40 & 87.52\%           & 18.56\% & 0.00\%  (9.8\%)  & 0.00\%  (10.6\%) & 0.00\% (1.7\%) \\
			%JumpReLU (R) & 0.40  & 88.63\%           & 13.79\% & 0.00\% (4.7\%)  & 0.00\% (5.7\%)  & 2.34\% (5.1\%) \\
			
			JumpReLU (R) & 0.50  & 88.20\%           & 20.13\% & 0.00\% (7.1\%)  & 0.00\% (7.9\%)  & 3.66\% (13.1\%) \\

			\midrule 
			\rowcolor{graytable}
			ReLU  (Robust) & 0.00  & 87.93\% & 51.88\%          & 0.00\%  (3.6\%)           & 0.00\% (4.2\%)           & 0.00\% (3.6\%)\\
			\rowcolor{graytable}
			JumpReLU (D)   & 0.40  & 86.19\% & 56.70\% & 0.00\% ({13.2\%}) & 0.00\% ({14.1\%}) & 0.00\% (4.3\%)  \\
			%\rowcolor{graytable}
			%JumpReLU (R)   & 0.40 & 87.21\% & \textbf{59.99\%}          & 0.00\% (\textbf{10.7\%})          & 0.00\% (\textbf{11.6\%})          & \textbf{18.74\%} (\textbf{11.52\%})\\
			
			\rowcolor{graytable}
			JumpReLU (R)  & 0.50  & 86.15\% & \textbf{61.03\%}         & 0.00\% (\textbf{13.6\%})          & \textbf{1.18\%} (\textbf{14.6\%})          & \textbf{15.79\%} (\textbf{19.7\%})\\

			\bottomrule 
	\end{tabular}}
	\caption{Results for AlexLike network (CIFAR10).}
    \label{tab:randrelu_alex}
	\end{subtable}

    \begin{subtable}{1\textwidth}	
	\centering
	\scalebox{0.99}{
		\begin{tabular}{lcccccccccc} \toprule
			Model      &   $\kappa$   & Accuracy  & PGD & DF$_\infty$ &DF$_2$ & TR\\
			\midrule
			ReLU  (Base) & 0.00  & \textbf{94.31\%}  & 0.37\% & 0.00\% (1.4\%)  & 0.00\% (1.8\%)  & 0.00\% (1.3\%) \\ 
			JumpReLU (D)  & 0.07 & 92.58\%           & 0.95\% & 0.00\% (14.3\%) & 0.00\% (18.5\%) & 0.00\% (1.9\%)  \\
			%JumpReLU (R)  & 0.07 & 93.38\%           & 9.84\% & 0.00\% (10.6\%) & 0.00\% (14.2\%) & 2.78\%(3.1\%) \\
			JumpReLU (R) & 0.09  & 92.53\%           &13.40\% & 0.00\% (14.7\%) & 0.00\% (18.3\%) & 7.53\% (5.9\%) \\

			\midrule 
			\rowcolor{graytable}
			ReLU  (Robust) & 0.00 & 93.72\%          & 60.43\%           & 0.00\% (6.4\%)           & 0.00\% (7.5\%)           & 0.00\% (4.8\%)\\                  
			\rowcolor{graytable}
			JumpReLU (D) & 0.07& 93.01\%  & 67.89\%        & 0.00\% ({44.4\%}) & 0.00\% (\textbf{43.8\%}) & 0.00\% (6.1\%)\\
			\rowcolor{graytable}
			%JumpReLU (R) & 0.07 & 93.46\%  & \textbf{69.66\%}      & 0.00\% (\textbf{34.4\%})          & 0.00\% (\textbf{36.3\%})          & \textbf{32.24\%} (\textbf{13.5\%})\\
			%\rowcolor{graytable}
			
			JumpReLU (R) & 0.09 & 93.07\%  & \textbf{71.82\%}           & 0.00\% (\textbf{44.5\%})     & 0.00\% (\textbf{43.8\%})          & \textbf{29.27\%} (\textbf{18.1\%})\\

			\bottomrule 
	\end{tabular}}
	\caption{Results for WideResNet (CIFAR10).}
    \label{tab:randrelu_wide}
	\end{subtable}

    \begin{subtable}{1\textwidth}	
	\centering
	\scalebox{0.99}{
		\begin{tabular}{lcccccccccc} \toprule
			Model    &   $\kappa$   & Accuracy & PGD  & DF$_\infty$  &DF$_2$  &TR\\
			\midrule
			ReLU  (Base) & 0.00 & \textbf{92.07\%} & 0.74\% & 0.00\% (0.7\%) & 0.00\% (0.9\%) & 0.00\% (0.7\%)\\
			
			JumpReLU (D) & 0.06 & 91.10\%          & 0.92\% & 0.00\% (5.3\%) & 0.00\% (6.8\%) & 0.00\% (1.0\%) \\
			
			%JumpReLU (R)& 0.06 & 91.16\%          & 3.46\% & 1.11\% (5.2\%) & 1.30\% (6.5\%) & 3.33\% (3.4\%)\\
			
			JumpReLU (R) & 0.08 & 90.37\%          & 5.07\% & 1.36\% (8.1\%) & 1.59\% (9.8\%) & 2.28\% (5.2\%)\\

			\midrule 
			\rowcolor{graytable}
			ReLU  (Robust) & 0.00 & 91.69\% & 53.98\%          & 0.00\% (4.7\%)           & 0.00\% (5.3\%)           & 0.00\% (4.1\%)\\                  
			\rowcolor{graytable}
			JumpReLU (D) & 0.06 & 90.12\% & 59.66\% & 0.00\% ({62.6\%}) & 0.00\% ({51.4\%}) & 0.00\% ({4.9\%})  \\
			\rowcolor{graytable}
			%JumpReLU (R) & 0.06 & 90.79\% & {65.29\%}          & 1.09\% (43.0\%)          &  1.24\% (39.7\%)          & \textbf{15.38\%} ({20.0\%})\\
			
			\rowcolor{graytable}
			JumpReLU (R) & 0.08 & 90.16\% & \textbf{68.98\%}          & \textbf{1.43\%} (\textbf{65.6\%})          & \textbf{1.68\%} (\textbf{53.3\%})          & \textbf{7.96\%} (\textbf{25.8\%})    \\            
			
			\bottomrule 
	\end{tabular}}
	\caption{Results for MobileNetV2 (CIFAR10).}
    \label{tab:randrelu_mobile}
	\end{subtable}
	
\end{table}

\section*{Acknowledgments}

We would like to acknowledge ARO, DARPA, NSF, and ONR for providing partial support for this work. 
We would also like to acknowledge Amazon for providing AWS credits for this project. 

{%\small
	\bibliographystyle{ieee}
	\bibliography{egbib}
}

\end{document}